\documentclass[11pt,preprint]{aastex}
\pdfoutput=1
\usepackage{bm}
\usepackage{graphicx}
\usepackage{amsmath}    
\usepackage{amssymb}    
\usepackage{equation}
\usepackage{footnote}
\usepackage{color}
\usepackage{amsmath}
\usepackage{mathtools}
\usepackage{pdflscape}
\usepackage{multirow}

\let\oldfootsep=\footnotesep
\newcommand\ltsima{$\; \buildrel <\over\sim \;$}
\newcommand\simlt{\lower.5ex\hbox{\ltsima}}
\newcommand\gtsima{$\; \buildrel >\over\sim \;$}
\newcommand\simgt{\lower.5ex\hbox{\gtsima}}
\newcommand\etal{et~al.}

\newcommand\msun {M_\odot}

\newcommand\mearth {{M_\oplus}}

\newcommand\pac{Paczy{\'n}ski }

\newcommand{\mathbold}[1]{\mbox{\boldmath $\bf#1$}}
\newcommand\piEbold{{\mathbold \pi_{\rm E}}}
\newcommand\mubold{{\mathbold \mu}}

\shorttitle{MOA-2008-BLG-379Lb Is a Super-Jupiter}
\shortauthors{Bennett et al}

\begin{document}


\title{Keck and Hubble Observations Show That MOA-2008-BLG-379Lb Is 
a Super-Jupiter Orbiting an M Dwarf}


\author{David~P.~Bennett\altaffilmark{1,2},
Aparna~Bhattacharya\altaffilmark{1,2},
Jean-Philippe~Beaulieu$^{3,4}$,
Naoki~Koshimoto$^{1,2,5}$,
Joshua~W.~Blackman\altaffilmark{6},
Ian~A.~Bond\altaffilmark{7},
Cl{\'e}ment~Ranc\altaffilmark{8},
Natalia~Rektsini\altaffilmark{3,4},
Sean K.~Terry\altaffilmark{1,2,9},
Aikaterini~Vandorou\altaffilmark{1,2},
Jessica~R.~Lu\altaffilmark{9},
Jean~Baptiste~Marquette\altaffilmark{10}, 
Greg~Olmschenk\altaffilmark{1,2},
and
Daisuke~Suzuki\altaffilmark{5}
 \\  } 
              

\affil{$^{1}$Code 667, NASA Goddard Space Flight Center, Greenbelt, MD 20771, USA;    \\ Email: {\tt bennettd@umd.edu}}
\affil{$^{2}$Department of Astronomy, University of Maryland, College Park, MD 20742, USA}
\affil{$^{3}$School of Physical Sciences, University of Tasmania, Private Bag 37 Hobart, Tasmania 7001 Australia}
\affil{$^{4}$Institut d'Astrophysique de Paris, 98 bis bd Arago, 75014 Paris, France}
\affil{$^{5}$Department of Earth and Space Science, Graduate School of Science, Osaka University, Toyonaka, Osaka 560-0043, Japan}
\affil{$^{6}$Physikalisches Institut, Universit\"{a}t Bern, Gessellschaftsstrasse 6, CH-3012 Bern, Switzerland}
\affil{$^{7}$Institute of Natural and Mathematical Sciences, Massey University, Auckland 0745, New Zealand}
\affil{$^{8}$Sorbonne Universitl{\'e}, CNRS, Institut d'Astrophysique de Paris, IAP, F-75014 Paris, France}
\affil{$^{9}$University of California Berkeley, Berkeley, CA}
\affil{$^{10}$Laboratoire d'astrophysique de Bordeaux, Univ. Bordeaux, CNRS, B18N, allée Geoffroy Saint-Hilaire, 33615 Pessac, France}


\begin{abstract}
We present high angular resolution imaging that detects the MOA-2008-BLG-379L exoplanet host star
using {\sl Keck} adaptive optics and the
{\sl Hubble Space Telescope}. These observations
reveal host star and planet masses of $M_{\rm host}=0.434\pm0.065 \msun$, and
$m_p=2.44 \pm 0.49 M_{\rm Jupiter}$. They are located at a distance of $D_L=3.44\pm0.53\,$kpc, with 
a projected separation of $2.70\pm 0.42\,$AU. These results contribute to
our determination of exoplanet host star masses for the \citet{suzuki16} statistical sample, which
will determine the dependence of the planet occurrence rate on the mass and distance of the host stars.
We also present a detailed discussion of the image constrained modeling version of the  \texttt{eesunhong} light curve
modeling code that applies high angular resolution image constraints to
the light curve modeling process.
This code increases modeling efficiency by a large factor by excluding models that are inconsistent with the high
angular resolution images.
The analysis of this and other events from the \citet{suzuki16} statistical sample
reveals the importance of including higher order effects, such as
microlensing parallax and planetary orbital motion even when these features are not required to fit
the light curve data. The inclusion of these
effects may be needed to obtain accurate estimates of the uncertainty of other microlensing parameters
that affect the inferred properties of exoplanet microlens systems.
This will be important for
the exoplanet microlensing survey of the {\sl Roman Space Telescope}, which 
will use both light curve photometry and high angular resolution imaging to characterize planetary
microlens systems.
\end{abstract}


\section{Introduction}
\label{sec-intro}
Gravitational microlensing surveys of the Galactic bulge have long been recognized as an effective
way \citep{mao91} to discover exoplanets down to low masses \citep{bennett96} orbiting beyond
the snow line \citep{gouldloeb92}. Because of this, the Astro2010 decadal survey recommended a
space-based exoplanet microlensing survey \citep{bennett02,bennett_MPF,WFIRST_AFTA,penny19}
to complete the statistical census of exoplanets in orbits $\simgt 1\,$AU to complement Kepler's survey
of planets in short period orbits \citep{kepler2018}.

Previous statistical studies of planetary microlensing events have revealed that super-Earths and
Neptunes are more common than higher mass planets \citep{sumi10,gould06,gould10,cassan12,suzuki16}.
The largest microlensing sample analyzed to date \citep{suzuki16,suzuki18} has revealed a contradiction
to a prediction based on the leading core accretion theory of planet formation \citep{lissauer_araa,pollack96}.
The standard core accretion theory includes a runaway gas accretion process, in which giant planet cores of
$\sim 10\mearth$ grow rapidly to masses similar to that of Jupiter ($318\mearth$) by accretion of Hydrogen and
Helium gas. This process led to predictions \citep{idalin04,mordasini09,emsenhuber21} of a sub-Saturn mass ``desert" in the
distribution of exoplanets, because it was thought to be very unlikely for gas accretion to terminate in the
middle of this rapid growth phase. However, the MOA (Microlensing Observations in Astrophysics) 
Collaboration microlensing results \citep{suzuki16} 
indicated a smooth, power-law distribution through this sub-Saturn mass region, in contradiction
to these earlier theoretical predictions \citep{suzuki18}, although more sophisticated theoretical calculations
do predict a power-law mass function down to mass ratios of $\sim 10^{-4}$ \citep{adams21}. In addition, a rigorous reanalysis of the 
\citet{mayor11} radial velocity exoplanet sample indicated no evidence for such a desert \citep{bennett21},
despite suggestions to the contrary in the \citet{mayor11} paper. The more recent radial velocity results from the
California Legacy Survey \citep{rosenthal21,fulton21} also show no evidence for such a sub-Saturn mass exoplanet
desert. These observations are consistent with three dimensional hydrodynamic simulations that show that 
the formation of a circumplanetary disk can slow gas accretion \citep{szulagyi14}, and the gas accretion
can also be slowed by collisions of protoplanets \citep{alidib22}. ALMA observations of gaps in protoplanetary disks
are also easier to explain \citep{nayakshin19,nayakshin22} with giant planet growth that is slower than
predicted by the runaway accretion scenario.

One characteristic of the exoplanets beyond the snow line found by microlensing that has not
been explored is the dependence of the planet occurrence rate, as a function of host mass, or more
simply, the planet hosting probability as a function of host mass. Kepler data has demonstrated a
dramatic difference in the planetary systems orbiting M-dwarfs and those orbiting more solar-like stars
of spectral types F, G, and K \citep{mulders15}. M dwarfs host many more small planets in short period
orbits than more massive host stars do. A different trend is expected for planets in wider orbits, beyond the
snow line. It is expected that gas giant planets will form more easily around more massive stars \citep{laughlin04}
and that the protoplanetary disks of M dwarfs will often lose their Hydrogen and Helium gas before large amounts
of gas can be accreted onto protoplanets. Microlensing has previously made mass measurements for 
two microlens planets with masses of $\sim 3 M_{\rm Jupiter}$ orbiting M dwarfs of mass $\sim 0.43 \msun$ 
have been reported \citep{poleski_ob120406,tsapras_ob120406,dong-ogle71,bennett_ogle71}, and in this paper,
we present mass measurements of the
MOA-2008-BLG-379Lb planet and host star masses very similar to these
previous examples. This suggests that the formation of super-Jupiter mass planets orbiting M dwarf hosts
is not as difficult as these theoretical predictions indicate, a feature that has also been seen in radial velocity
studies \citep{schlecker22}. Perhaps, this is not surprising given that the
\citet{laughlin04} prediction is based on the runaway gas accretion process that also predicted the
sub-Saturn mass ``desert\rlap", which is contradicted by the microlensing and radial velocity data.
A statistical analysis of a sample of events with mass measurements is necessary for a definitive tests of these predictions, 
and both the MOA-2008-BLG-379Lb and OGLE-2005-BLG-071Lb planets are part of the \citet{suzuki16} 
statistical sample that we are obtaining mass measurements for. However, 
preliminary analyses of both this \citet{suzuki16} microlensing
sample and the California Legacy Survey radial velocity sample \citep{rosenthal21,fulton21} indicate that 
the exoplanet hosting probability for wide orbit planets scales as roughly the first power of the host star
mass\footnote{https://clementranc.github.io/microlensing25/schedule/talks/bendav.html}.

It is also thought that wide orbit planets ranging in mass from $\sim 10\mearth$ to a few Jupiter masses 
may be needed \citep{raymond04,raymond07,childs22}
to create habitable conditions on terrestrial planets in inner orbits. These wide orbit planets are expected to be crucial 
for the delivery of water and other ingredients that may be needed for life to develop on these potentially 
habitable planets \citep{grazier16,osinski20,sinclair20}. Thus, an understanding of the host mass dependence
of planets found by microlensing may help us gain an understanding of which planetary systems might include
habitable planets. The planetary system we study in this paper has a mass ratio of $\sim 5\times 10^{-3}$, which
would indicate a super-Jupiter mass planet orbiting an M dwarf if the host mass is in the range 
$0.19 < M_{\rm host}/\msun < 0.6$, which is, in fact, what we find.

In this paper, we use adaptive optics (AO) observations with the NIRC2 instrument on the {\sl Keck}-2 telescope
and {\sl Hubble Space Telescope} observations
to identify the lens and planetary host star and provide a precise measurement of the masses and
distance of the MOA-2008-BLG-379L planetary system.

This paper is organized as follows. In Section~\ref{sec-event} we present the light curve of
microlensing event MOA-2008-BLG-379 and explain the challenges posed by the faintness of
its source star. In Section~\ref{sec-HST_Keck}, we describe the Hubble Space Telescope and {\sl Keck} high angular 
resolution follow-up observations and their analysis. Section~\ref{sec-LC} presents a new
method to apply the constraints from high angular resolution to the light curve modeling analysis.
The constraints from the light curve models and high angular resolution follow-up observations are
combined with relatively weak constraints from a Galactic model to derive the physical properties
of the MOA-2008-BLG-379Lb planetary system in Section~\ref{sec-lens_prop}, and then in section~\ref{sec-RGES_lessons}
we describe the implications of our high angular resolution imaging program for the Roman Galactic Exoplanet Survey.
We discuss the implications of these
results and present our conclusions in Section~\ref{sec-conclude}, and Appendix~\ref{sec-old-lc} compares the
light curve model presented by \citet{suzuki14,suzuki14e} with models using the \texttt{eesunhong} code and with improved
MOA photometry.

\section{Microlensing Event MOA-2008-BLG-379}
\label{sec-event}

\begin{figure}
\epsscale{0.9}
\plotone{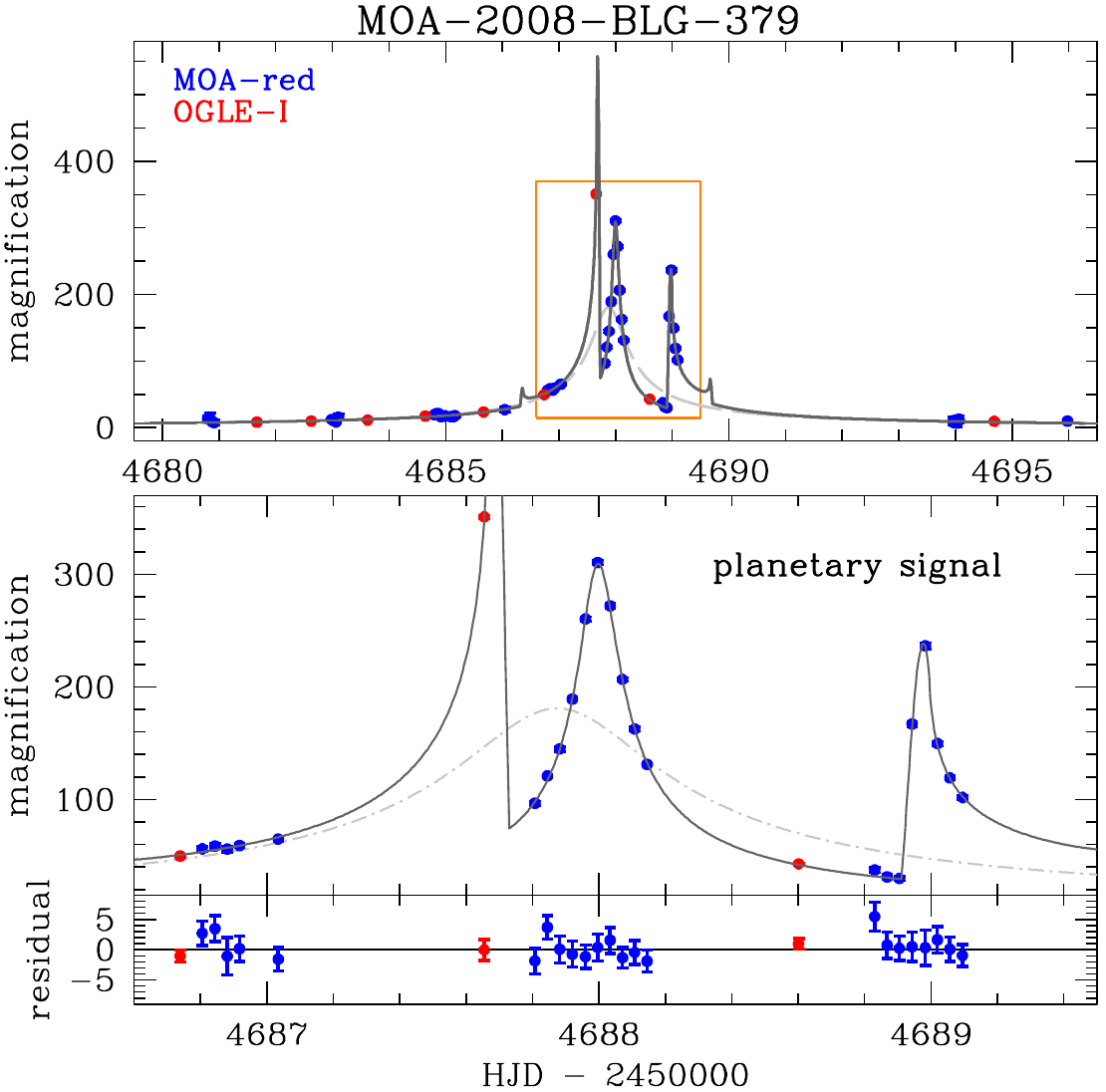}
\caption{The best fit light curve with the constraints from the high angular resolution follow-up data,
as explained in Section~\ref{sec-LC}.  This is the model from the third column of 
Table~\ref{tab-Cmparams}, with $u_0 > 0$ and $s > 1$.}
\label{fig-lc}
\end{figure}

MOA-2008-BLG-379 \citep{suzuki14,suzuki14e}, located at 
$\rm(R.A., decl.)(J2000)$ = ($17^{\rm h}58^{\rm m}49^{\rm s}.44$, -$30^{\circ}11'48''.95$), 
is an unusual planetary microlensing event in that despite a very strong planetary
signal, it was not identified as a planetary microlensing event until several years after it was observed,
even though the light curve photometry from the MOA and OGLE groups could be viewed on
the microlensing alert websites very shortly after the data were taken. The event was first discovered
by the MOA alert system \citep{bond01} at UT 22:00, 2008 August 9, (or HJD' = HJD-2450000 = 4688.42).
The data taken on the next night revealed a strong caustic entry feature that make it clear that this
was not a single lens microlensing event, and data from the OGLE group soon confirmed this conclusion
when they independently discovered the event two weeks later. (The OGLE discovery was delayed because
it was found during the development of the ``new object" channel of the OGLE Early Warning System \citep{ogle-ews} 
that could find microlensing events, like MOA-2008-BLG-379, with faint source stars that were 
not close to an apparent ``star" identified in the OGLE reference image.)

While the light curve, shown in Figure~\ref{fig-lc}, was clearly not that of a single lens event, it was 
observed at a time when only seven 
planetary microlensing events had been published 
\citep{bond04,udalski05,ogle390,gould06,gaudi-ogle109,bennett08}, and three more were 
well known to the microlensing community while under analysis
\citep{dong-moa400,sumi10,bennett16}. MOA-2008-BLG-379 showed dramatic deviations from a single lens 
light curve over about half of its apparent duration, and it was not immediately recognized that an event like
this could be due to very high magnification by a planetary lens system with a very faint source star.
The faintness of the source star meant that much of the lower magnification part of
the light curve did not rise above the photometric noise. 

The planetary nature of this event was discovered several years later as a part of the statistical analysis
that led to the MOA Collaboration study of exoplanet demographics beyond the snow line \citep{suzuki16}.
It is now generally understood that a light curve like the one shown in Figure~\ref{fig-lc} can only
be explained by a microlensing model with a planetary mass ratio below the International Astronomical
Uniion (IAU) preferred mass ratio threshold of 
of $q < 0.0400642$ \citep{exoplanet_IAU}.

While the planetary nature of the light curve is clear, there is an additional complication due to the faint source
star. This is due to the formula for microlensing magnification by a single, compact lens, 
$A = (u^2 + 2)/(u\sqrt{u^2+4})$, where $u$ is the lens-source separation in units of the Einstein radius.
As a single lens approaches high magnification, the lens-source alignment becomes nearly perfect, so that
$u\rightarrow 0$. In this limit, we have $A\approx 1/u$, 
so that the apparent brightness becomes $F_{s,{\rm obs}}, \simeq F_s/u$ where $F_{s,{\rm obs}}$ and
$F_s$ are the magnified and unlensed brightnesses of the background source, respectively. This is a difficulty because main 
sequence stars are not 
individually resolved in ground based observations of the crowded Galactic bulge fields, where most
gravitational microlensing events are observed. Thus, the brightness of the source star is normally determined
from the light curve model fit, but this can be a problem for high magnification microlensing events. When the 
high magnification approximation, $F_{s,{\rm obs}}, \simeq F_s/u$, applies, the light curve shape will only reveal the combination
$F_s/u$, but not $F_s$ and $u$ individually. 
Thus, $F_s$ can only be determined from the lower magnification parts of the light curve, where the 
high magnification approximation does not apply. When the source star is faint, this becomes difficult
and sensitive to low-level light curve systematic errors. So, the faintness
of the MOA-2008-BLG-379 source implies that the light curve modeling will result in imprecise measurements
of $t_{\rm E}$ and $F_s$. Both of these parameters are used in the mass and distance determination for the
lens system, so this is an important
part of the rationale for the new modeling code that we discuss in  
Section~\ref{sec-LC}. This new code applies the constraints from the high angular resolution images
to the light curve models.
 
We have used an improved data reduction method using the difference imaging code of \citet{bond01}, but 
we have applied the detrending method of \citet{bond17} to remove systematic errors, including the 
color dependent effects of differential refraction \citep{bennett12} that are enhanced by the wide MOA-red 
passband used for the MOA-II survey. This method also calibrates the MOA data to the OGLE-III
catalog \citep{ogle3-phot}.
However, since no MOA $V$-band data was taken in 2008, we must use a $V-I$ color from 
other observations to relate the MOA-red magnitude to the $I$-band. We find
\begin{equation}
I = R_{\rm MOA} + 28.1940 - 0.2002 (V-I) \ ,
\end{equation}
where the MOA-red magnitude, $R_{\rm MOA}$, is related to the MOA instrumental flux units, $F_{\rm MOA}$, by
$R_{\rm MOA} = -2.5\log_{10}(F_{\rm MOA}) $. We will use the {\sl Hubble} observations to help determine the 
$V-I$ color of the source star.

\section{Hubble Space Telescope and Keck Follow up Observations and Analysis}
\label{sec-HST_Keck}

The first high angular resolution follow-up observations for MOA-2008-BLG-379 were taken by {\sl Hubble}  on
2013 October 9, which was shortly
after the planetary nature of the event was discovered and before the planetary discovery paper
was published \citep{suzuki14,suzuki14e}. {\sl Hubble} Space Telescope program GO-12541 had already
been approved for two epochs of follow-up observations of four other planetary microlensing events. However,
the first epoch observations of OGLE-2005-BLG-169 were sufficient to determine the physical parameters of 
this event \citep{bennett15}, particularly when combined with a later epoch of {\sl Keck} Telescope adaptive optics 
imaging \citep{batista15}. So, we were able to switch the target for these second epoch observations from 
OGLE-2005-BLG-169 to the event we analyze in this paper, MOA-2008-BLG-379. We obtained 16 {\sl Hubble}
Wide Field Camera 3 UVIS images of this event with $8\times 70\,$sec and $8\times 125\,$sec dithered exposures 
in the F814W and F555W passbands, respectively. However, the 
MOA-2008-BLG-379 source star is a magnitude fainter than the OGLE-2005-BLG-169, so we probably would have
asked for observations using two instead of one  {\sl Hubble} orbit if we had originally proposed to
observe the MOA-2008-BLG-379 event. As a result, the S/N of the {\sl Hubble} data was lower than desired, and 
our initial analysis did not separately detect the source and lens stars.

The data used in this paper can be found in MAST using the following DOIs. The {\sl Hubble} are at: 
\dataset[10.17909/edn8-2564]{http://dx.doi.org/10.17909/edn8-2564},
and the {\sl Keck} data for the May, 2018 observations can be found in 
\dataset[10.26135/KOA3]{http://dx.doi.org/10.26135/KOA3}, and
\dataset[10.17909/edn8-2564]{http://dx.doi.org/10.17909/edn8-2564} contains the 
August, 2018 data. Note that DOIs for the Keck observations link to observations of many events. Filter for the
observations used in this paper with the target name mb08379.

\subsection{Keck Data Analysis}
\label{sec-Keck}

Because of this, we also observed this event in 2018 as a part of our NASA {\sl Keck} Key Strategic Mission Support (KSMS) 
``Development of the WFIRST Exoplanet Mass Measurement Method\rlap," using the laser guide star adaptive 
optics mode of the NIRC2 instrument with the $K_s$ filter on the {\sl Keck}-2 telescope. 
In order to calibrate the {\sl Keck} photometry, we obtained 10 images with 30 second exposures  
using the NIRC2 wide camera on 27 May 2018. The wide camera images cover a $1024\times 1024$ pixel
area with a plate scale of $39.686\,$mas per pixel. We adopted a 5 points dither pattern with a step of 2 arcsec 
for the 10 images. These $K_s$ wide camera images were flat field and dark current corrected using standard
methods \citep{batista14,beaulieu16}, and then, we performed the sky correction 
and stacked the images using the SWarp Astrometics package \citep{SWarp}. We used the SExtractor \citep{sextractor}
package to obtain photometry with a 10 pixel aperture.  We cross-identified the detected sources with our re-reduction 
of $K$ band images from the Vista VVV survey \citep{minniti-vvv} as described by \citet{beaulieu_vvv}. We select 
39 cross-identified stars to obtain the zero point for the {\sl Keck} photometry. We then obtained the measured flux of the 
combined lens plus source (or stars 1$+$2) images to be $K_{s12} = 18.09\pm 0.05$, with a calibration error of 2\%.

The detailed analysis of the blended background source star and foreground lens (and planetary host) star
requires higher angular resolution than the NIRC2 wide camera provides, so the NIRC2 narrow camera
was used. The NIRC2 narrow camera uses the same $1024\times 1024$ pixel detector as the wide camera,
but the plate scale is $4\times$ smaller at $9.942\,$mas per pixel. The first set of NIRC2 narrow camera images of our
target were taken on 26 May 2018 with a small dither pattern and each of 18 frames consisting of two co-added 30 second
exposures. Another set of 17 fames, each consisting of three co-added 20 second exposures where taken 
on 6 August 2018, with a similar small dither pattern. All images were taken with the $K_s$ filter.

\begin{figure}
\epsscale{0.9}
\plotone{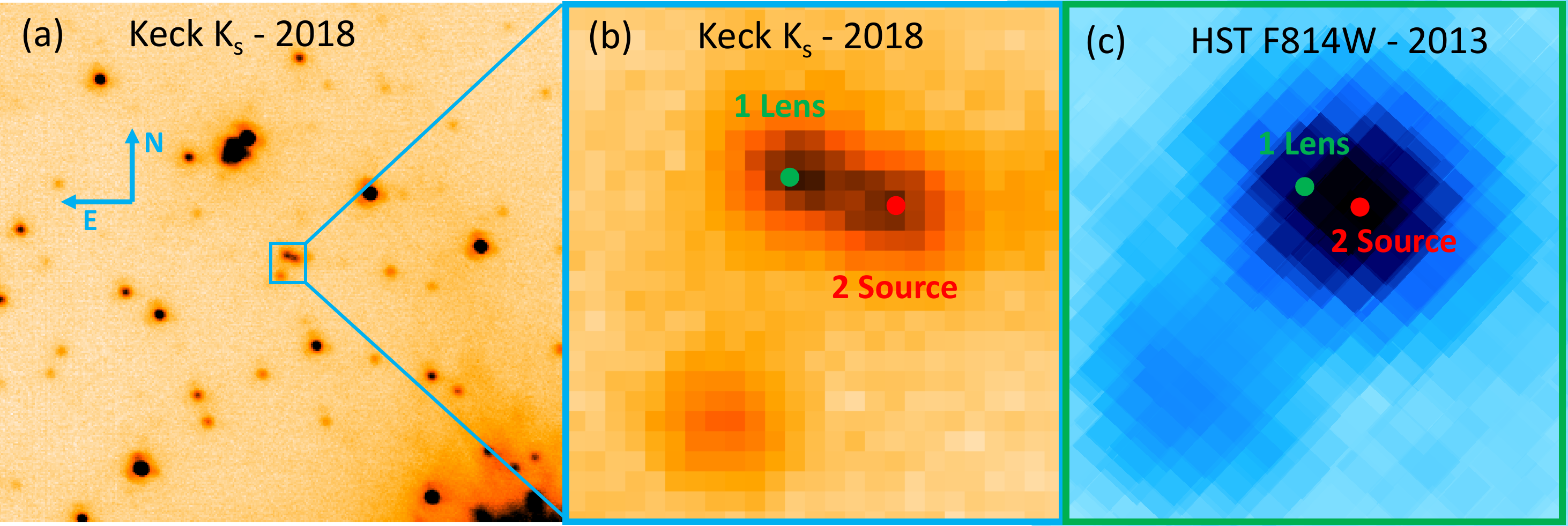}
\caption{(a) A $32^{\prime\prime}\times 32^{\prime\prime}$ section of the
coadded sum of 12$\times$60-sec exposures with the {\sl Keck}-NIRC2 narrow camera images taken
10 years after the microlensing event, with the target location indicated by a blue square. (b) and (c) are 
$2.5^{\prime\prime}\times 2.6^{\prime\prime}$ close-ups of the target with the
{\sl Keck} $K_s$ band and the {\sl Hubble} WFC3/UVIS F814W passband, respectively. The green and red dots are 
the best fit positions of the lens and source stars.
The co-added {\sl Hubble} frame, taken in 2013, involves no image resampling. Instead, 
each pixel of the individual {\sl Hubble} images is divided into $100\times 100$ sub-pixels which are each 
assigned the same flux value. Thus, the image (c) shows full sized pixels with the observed dither offsets accurate to $0.01\,$pixels.
\label{fig-Keck}}
\end{figure}

The images taken in May and August were analyzed separately. The raw images were flat fielded and bias subtracted, and
then bad pixels and cosmic rays were removed from the raw images. Then these cleaned, raw images were corrected for
geometric distortion, differential atmospheric refraction, and then stacked into a single co-added frame for each of the 
May and August data sets using the methods of  \citet{jlu_thesis,KAI}. This process also resulted in the removal of
a few lower quality images from each of the May and August data sets. The images included in these final stacked images
had mean and RMS PSF full-width, half-max (FWHM) values of ${\rm FWHM} = 71.5\pm 2.8\,$mas for the 13 good images from the
May data set and ${\rm FWHM} = 58.6\pm 4.4\,$mas for the 12 good images from the
August data set. Figure~\ref{fig-Keck}(a) shows the stacked NIRC2 narrow camera image from August 2018, and 
Figure~\ref{fig-Keck}(b) shows a close-up image, approximately $250\,$mas on a side at the location of the 
MOA-2008-BLG-379 microlensing event. This event involved stars 1 and 2, which we will identify as the 
lens and source stars with the help of the earlier {\sl Hubble} images.

We have analyzed these co-added NIRC2 narrow camera images following the
method of \citet{aparna18} using DAOPHOT \citep{Daophot}.
The DAOPHOT PSF models were built in two stages. First, we ran DAOPHOT's FIND and PHOT commands
to find all the stars in the image, and then we used the DAOPHOT PICK command to build a list of 
bright ($K_s < 18.5$) isolated stars that can be used to construct our empirical PSF models. Our target
``star" is actually bright enough to pass this magnitude cut, but by 2018 the images of the two stars had separated 
enough to be resolved, and both stars were fainter than $K_s = 18.5$. However, our analysis must always exclude the 
target from the candidate PSF star list because it must necessarily consist of the lens and source stars separating 
after the microlensing event.
The PSF stars were selected to
be located close to the target with a roughly even distribution on all sides of the target to minimize any effects
of a spatially varying PSF.

Once we have built the PSF model for each co-added frame, we use DAOPHOT's ALLSTAR routine to fit all the stars
in the image, including the lens and source star pair that are shown in panels (a) and (b) of Figure~\ref{fig-Keck}. 
The location of the source and lens stars were determined using the standard method using difference images
taken near peak magnification \citep{bennett06,bennett-ogle109,sumi10}, which is then transformed to the 
NIRC2 wide camera image, which has been astrometricly matched to the coordinate system of the MOA
reference image.

\begin{deluxetable}{cccc}
\tablecaption{Keck NIRC2 Narrow Camera Fit Parameters \label{tab-Keck}}
\tablewidth{0pt}
\tablehead{Data Set & $K_{s1} - K_{s2}$ & ${\rm RA}_1 - {\rm RA}_2\,$(mas) &  ${\rm DEC}_1 - {\rm DEC}_2\,$(mas)} 
\startdata
2018 May & $-0.2058 \pm 0.0661$ & $55.91 \pm 1.55$  & $18.57 \pm 0.80$ \\
2018 Aug & $-0.0325 \pm 0.0264$  & $53.79 \pm 0.69$  & $19.44 \pm 0.44$ \\
weighted sum & $-0.0564 \pm 0.0245$  & $54.14 \pm 0.63$  & $19.24 \pm 0.38$ \\
\tableline
\tableline
\vspace{0.1cm}
 & & $\mu_{\rm rel,H,E}$(mas/yr) & $\mu_{\rm rel,H,N}$(mas/yr)  \\
\tableline
 & & $5.430 \pm 0.063$  &  $1.926 \pm 0.038$ \\
\enddata\\
\vspace{0.15cm}
\parbox{13.3cm}{
The separation was measured by {\sl Keck} $K$-band images taken 9.9900 years after the peak of the event. The follow-up observation was taken on Aug 06, 2018 and the peak of the microlensing light curve event 
was on August 9, 2008. }
\end{deluxetable}

The error bars are determined with the jackknife method \citep{quenouille1949,quenouille1956, tukey1958, tierney}
following \citet{aparna21}. This method is able to determine uncertainties due to the PSF variations in
the individual images. For the May data, the jackknife method requires that $N = 13$ different ``jackknife" co-added images 
are constructed, with each of the 13 good May images being excluded from one of these jackknife images.
These images are all then analyzed with same dual star PSF models as the full combined image of all 13 good images, yielding
13 sets of dual star fit parameters. We use the mean of the these the parameters in these jackknife reduction
as our best fit parameters, shown in Table~\ref{tab-Keck}, with the uncertainty for a parameter, $x$, given by the
jackknife formula,
\begin{equation}
\label{eq-jackknife}
\sigma_x = \sqrt{\frac{N-1}{N} \sum (x_{i} - \bar{x})^{2}} \ ,
\end{equation}  
where $x_{i}$ is the parameter value from the $i$th jackknife image stack and $\bar{x}$ is the mean value
of the parameter from the jackknife images. This Equation \ref{eq-jackknife} is the same 
formula as the sample mean error, except that it is multiplied by $\sqrt{N-1}$ to account for the fact that
each individual image is included in all, but one of the jackknife image stacks. The August data are reduced
in the same way, except that the number of good images is $N = 12$

The results of our analysis are summarized in Table~\ref{tab-Keck}, which shows the magnitude difference,
and separation measurements from the May and August {\sl Keck} reductions. The reported values are the mean
of the measurements from the jackknife runs, and the error bars are determined from equation~\ref{eq-jackknife}.
The August data have a smaller scatter than the May data due to somewhat distorted PSF shapes in the May data.
The third line of Table~\ref{tab-Keck} shows the weighted sum of these measurements, and it appears that the 
May measurement of the magnitude difference is more than 2$\sigma$ larger than the 
weighted sum of the magnitude difference. However, we should note that there are only 13 images that contribute to
the jackknife error bars. So, the precision of the error bar estimates is subject to a Poisson uncertainty
of $\sim 1/\sqrt(13) = 28$\%. Thus, a 1$\sigma$ increase in the error bar would bring the magnitude difference
from May to within 2$\sigma$ of the weighted average value. We will use the weighted sum values
for the remainder of our analysis. The magnitude of the lens-source relative proper motion in the Heliocentric
frame is $\mu_{\rm rel,H} = 5.761\pm 0.061\,$mas/yr.

Since the magnitude difference of the two stars is only $K_{s1}-K_{s2} = -0.0564 \pm 0.0245$, we cannot
use the estimated $K_s$ magnitude of the source star from the light curve model to determine which star
is the source star to high confidence. This is why we label the stars with numbers 1 and 2, in addition to the lens 
and the source. While the {\sl Keck} observations do not allow us to determine which of stars 1 and 2 is the source
star, we will see in Section~\ref{sec-hst} that our earlier, 2013, {\sl Hubble} observations will answer this question.
With the calibration of the combined stars 1$+$2 flux from the NIRC2 wide camera, we have $K_{s1} = 18.815 \pm 0.056$
and $K_{s2} = 18.871 \pm 0.056$.



\subsection{Hubble Data Analysis}
\label{sec-hst}

As mentioned in the introduction to Section~\ref{sec-HST_Keck}, we obtained a single orbit of {\sl Hubble} observations
from program GO-13417, using
the WFC3/UVIS camera, in 2013, five years after the event. We obtained $8\times 70\,$sec.\ dithered exposures with the F814W
filter and $8\times 70\,$sec.\ dithered exposures with the F555W filter using the UVIS2-C1K1C-SUB aperture to minimize CTE
losses and minimize readout times in order to obtain 16 dithered images in a single orbit.
(The {\sl Hubble} data used in this paper can be found in MAST: \dataset[10.17909/edn8-2564]{http://dx.doi.org/10.17909/edn8-2564}.)
The analysis was done with a modified version of the codes used by
\citet{bennett15} and \citet{aparna18}, and these codes analyze the data from the original images without any resampling
in order to avoid the loss of resolution that the combination of dithered, undersampled images would provide.
Figure~\ref{fig-Keck}(c) shows a close-up of the 8 dithered F814W images registered to the same physical coordinate
system, plotted on top of each other. This is a representation of the data that our analysis code uses, because we 
simultaneously analyze the 8 individual images with pixel positions transformed to the same physical coordinate 
system. Because the {\sl Hubble} and {\sl Keck} images were taken 5.1657 and 9.9900 years after the event, respectively,
the {\sl Hubble}  images should show a lens-source separation that is about a factor of two smaller than the
separation seen in the {\sl Keck} image. The angular resolution of the {\sl Hubble}  images is also worse than the 
angular resolution of the {\sl Keck} images because of {\sl Hubble}'s smaller aperture, and relatively large,
undersampled pixels. (However, the much more stable PSF shapes delivered by {\sl Hubble} help to compensate 
for the lower angular resolution in this type of analysis.)

With the lower angular resolution of the  {\sl Hubble}  images, we had some concern that the image of the
fainter star to the South-West of the lens and source might interfere with the measurement of the lens-source
separation, so we have included this third star in our PSF fitting procedure. Also, the {\sl Keck} data provides 
a higher S/N measurement of the 2-dimensional separation between the lens and source stars, so we have 
added the option of applying a constraint to the two dimensional separation of stars 1 and 2 in our
three star {\sl Hubble} PSF fitting code. Because the {\sl Hubble} images were obtained earlier than the {\sl Keck} images, the separation of
the lens and source star should be $0.51709\times$ smaller in the {\sl Hubble} images than in the {\sl Keck} images.

The coordinate transformation between the {\sl Keck} and  {\sl Hubble}  images was done with 17 stars brighter than
$K_s < 14.8$, yielding the transformation 
\begin{equation}
\begin{eqalign}
         x_{\rm hst} =& 0.184169\, x_{\rm keck}-0.171462\, y_{\rm Keck}+483.4065\\
         y_{\rm hst} =& 0.171174\, x_{\rm keck}+0.183673\, y_{\rm Keck}+399.7104 \ ,
\label{eq-keck2hst}
\end{eqalign}
\end{equation}
from {\sl Keck} to {\sl Hubble} WFC3/UVIS pixels. The RMS scatter for this relation is 
$\sigma_x = 0.33$ and $\sigma_y = 0.28$ WFC3/UVIS pixels for the 17 stars used for the transformation. The {\sl Keck}
images were taken 5 years after the  {\sl Hubble} images, and the WFC3/UVIS pixels subtend 40\,mas. So, the
$\sim 12\,$mas scatter in the $x$ and $y$ coordinates could be fully explained by an average proper motion of
$2.4\,$mas per year in each direction. This magnitude of proper motion is typical of bulge stars, so it seems likely
that the scatter is largely explained by stellar proper motion of the astrometric reference stars.

\begin{deluxetable}{cccccccc}
\tablecaption{Hubble Multi-star PSF Fit Results \label{tab-hst}}
\tablewidth{0pt}
\tablehead{Filter & $\mubold_{\rm rel,H}$ & $x_{\rm hst1}$(pix) & $y_{\rm hst1}$(pix) & $x_{\rm hst2}$(pix) & $y_{\rm hst2}$(pix) & $F_{\rm hst1}$ & $F_{\rm hst2}$ \\
 & const. & \multicolumn{2}{c} {star\,$1 =$\,lens} & \multicolumn{2}{c} {star\,$2=$\,source} & (lens) & (source) } 
\startdata
{\bf F814W} & {\bf yes} & $\mathbold{-0.513(78)}$ & $\mathbold{-0.206(42)}$ & $\mathbold{0.171(74)}$ &  $\mathbold{0.100(73)}$ & $\mathbold{316(133)}$ & $\mathbold{941(128)}$ \\
F814W & no & $-0.389(116)$ & $-0.242(84)$ &  $0.168(60)$  & $0.139(47)$ & $389(136)$ & $867(135)$  \\
{\bf F555W} & {\bf yes} & $\mathbold{-0.524(47)}$ & $\mathbold{-0.235(37)}$ & $\mathbold{0.164(74)}$ &  $\mathbold{0.061(28)}$ & $\mathbold{50(42)}$ & $\mathbold{680(38)}$  \\
F555W & yes &   $0.769(43)$ &  $0.329(32)$ & $0.081(37)$ & $0.028(24)$ & $55(38)$  & $677(33)$  \\
\tableline
\enddata\\
\vspace{0.15cm}
\parbox{16.1cm}{
The coordinate system used here is 
centered on the center of the blended image of stars 1 and 2 in a preliminary reduction of an
F814W that was arbitrarily selected as the reference frame.}
\end{deluxetable}

Equation~\ref{eq-keck2hst} allows us to convert the {\sl Keck} relative proper motion values ($\mubold_{\rm rel,H}$) given at the 
bottom of Table~\ref{tab-Keck} to constraints on the positions of the source and lens stars in the {\sl Hubble}
images, taking into account the 5.1657 year interval between the microlensing event peak and the  {\sl Hubble}
observations. Table~\ref{tab-hst} shows the the positions and instrumental fluxes of the two stars of interest from our
constrained and unconstrained 3-star PSF fitting procedure. (The third star is included in the fit to prevent it 
from biasing the measurements of the lens and source stars.)
 As we shall see below, these results allow us to identify
star 1 and the lens (and planetary host) star and star 2 as the source star. The instrumental fluxes of stars 1 and 2
in Figure~\ref{fig-Keck}(c) are denoted by
 $F_{\rm hst1}$ and $F_{\rm hst2}$, and the 2 or 3-digit numbers given in parentheses are the 
uncertainties of the last 2 or 3 decimal places for each measurement.
We have analyzed the {\sl Hubble} F814W data both with and without this proper motion constraint,
but the F555W images can detect the fainter star at only 1$\sigma$ precision, so we have only done constrained 
fits for this passband. The rows shown in boldface are the ones used for our final analysis.

The F814W fits converged to a unique solution with star 2, to the South-East as the
brighter star, but the F555W fits with a constraint on $\mubold_{\rm rel,H}$ were fit almost equally well with 
star 1 or star 2 being the brighter star, which is not a surprise, since the fainter star is $<1.5\sigma$ from zero flux.
Our reduction code puts the F814W and F555W coordinates in the same reference frame, so each star should have
positions that are consistent between the two passbands.
The F555W model highlighted in boldface gives positions for stars 1 and 2 that are consistent with the positions
listed in the first F814W row (also highlighted in boldface). The measurements from these rows can be averaged to find
the weighted mean positions, and this yields average positions for both stars of $\chi^2 = 1.80$ for the 8 measurements 
(2 coordinates for each star in each passband), 4 parameters (the mean $x$ and $y$ values for each star), with two constraints 
(the separations implied by the $\mubold_{\rm rel,H}$ measurement). Thus, we have $\chi^2/{\rm dof} = 0.30$. 
In the F555W model listed in the bottom row, we also label the brighter star to be star 2, but now star 1 is located in the 
opposite direction - to the North-West of star 2. The star 2 position is still marginally consistent with the
F814W star 2 position, but the star 1 positions are pretty far from each other. The fit to average star 2
position using this bottom gives $\chi^2 = 3.83$, but the fit for the mean star 1 position gives $\chi^2 = 308.82$.
So, we have rejected this alternative F555W PSF fit model.

%

The {\sl Hubble} data were calibrated to the OGLE-III catalog \citep{ogle3-phot} using 7 relatively bright OGLE-III stars
that were matched to isolated stars in the {\sl Hubble} catalog. These calibrations give $I_2 = 21.56\pm 0.15$, 
$V_2 = 23.67 \pm 0.06$, $I_1 = 22.75 \pm 0.49$, and $V_1 = 26.49^{+1.93}_{-0.66}$. As indicated in
Table~\ref{tab-hst}, the $V$-band (F555W) brightness of star 1 is very marginally detected at $\sim 1\sigma$ significance.
The relatively large $I$-band uncertainties are largely due to the small lens-source separation of $\sim 0.74\,$pixels, 
which allows flux to be traded between the two stars \citep{bennett07}. As a result, the magnitude of both the lens and 
source stars combined is measured with higher precision. We find the magnitude of the combination of
stars 1 and 2 is $I_{12} = 21.250\pm 0.011$.

With these $V$ and $I$-band measurements, we can now determine which star is the source and which is the lens.
The discovery paper \citep{suzuki14,suzuki14e} determined the source star $I$ magnitude to be $I_S = 21.30\pm 0.03$ with
a color of $V_S-I_S = 2.29\pm 0.14$. However, since that analysis, the MOA group has begun detrending its photometry
to remove systematic errors caused by the apparent motion of nearby stars of different colors due to atmospheric refraction.
We used the detrending method of \citet{bond17} to correct this data, and following \citet{suzuki14,suzuki14e},
excluded the data points that obtained prior to March 17, 2008 and after October 22, 2008. This analysis yielded 
a best fit source magnitude of $I_S = 21.40$, which is just over $1\sigma$ brighter than the {\sl Hubble} $I$-band magnitude for star 2
and is much brighter than {\sl Hubble} $I$ brightness of star 1. However,
the detrending method of \citet{bennett12}, which is less aggressive at removing trends due to variations in seeing, 
yielded predicted source brightnesses of $I_S \simlt 21.0$. The best fit models with longer durations of MOA data
yielded even brighter source stars, and the exclusion of baseline observations with high airmass and poor seeing
did not bring the best fit source magnitude any closer to the {\sl Hubble} values. This 
uncertainty in the source brightness is due to the fact that it is only the low-magnification part of the light curve that
constrains the Einstein radius crossing time, $t_{\rm E}$, and the source brightness. Thus, high magnification events with faint
sources, like MOA-2008-BLG-379, are susceptible to low level systematic errors that can perturb the correct $t_{\rm E}$
and $I_S$ values. This is sometimes referred to as the blending degeneracy \citep{alard97,distefano95}.
Nevertheless, the light curve data, clearly favor the identification of star 2 as the source star. In contrast, the $I_1 = 22.75 \pm 0.49$ 
magnitude is considerably fainter than the light curve models predict.

The color of star 2, $V_2 -I_2 = 2.11\pm 0.16$ also matches the \citet{suzuki14,suzuki14e} color prediction of $V_S-I_S = 2.29\pm 0.14$,
and this color measurement is not affected by the blending degeneracy. The measured $V_1$ magnitude is quite uncertain,
since the detection of this star is very marginal in the $V$-band. The best fit color for star 1 is $V_1 - I_1 = 3.74$, and even if
we take the $2\sigma$ upper limit on the star 1 $V$-band brightness from Table~\ref{tab-hst}, we have $V_1 - I_1 = 2.68$,
which is still considerably redder than the source color from the light curve models. So, we identify star 2 to be the source star
and star 1 to be the lens and planetary host star, as we have labeled in Figure~\ref{fig-Keck}. With the identification of star 1 
with the lens star, the direction of motion of the lens
star with respect to the source star is $\sim 40^\circ$ from the direction rotation of the Galactic disk. Since the disk rotation
is a substantial fraction of the total velocity difference between disk and bulge stars, this direction of relative proper motion
is much more likely for lens stars in the disk (assuming a bulge star source) than the $\sim -140^\circ$ angle that would be implied
if star 2 was the lens star.

\begin{figure}
\epsscale{0.7}
\plotone{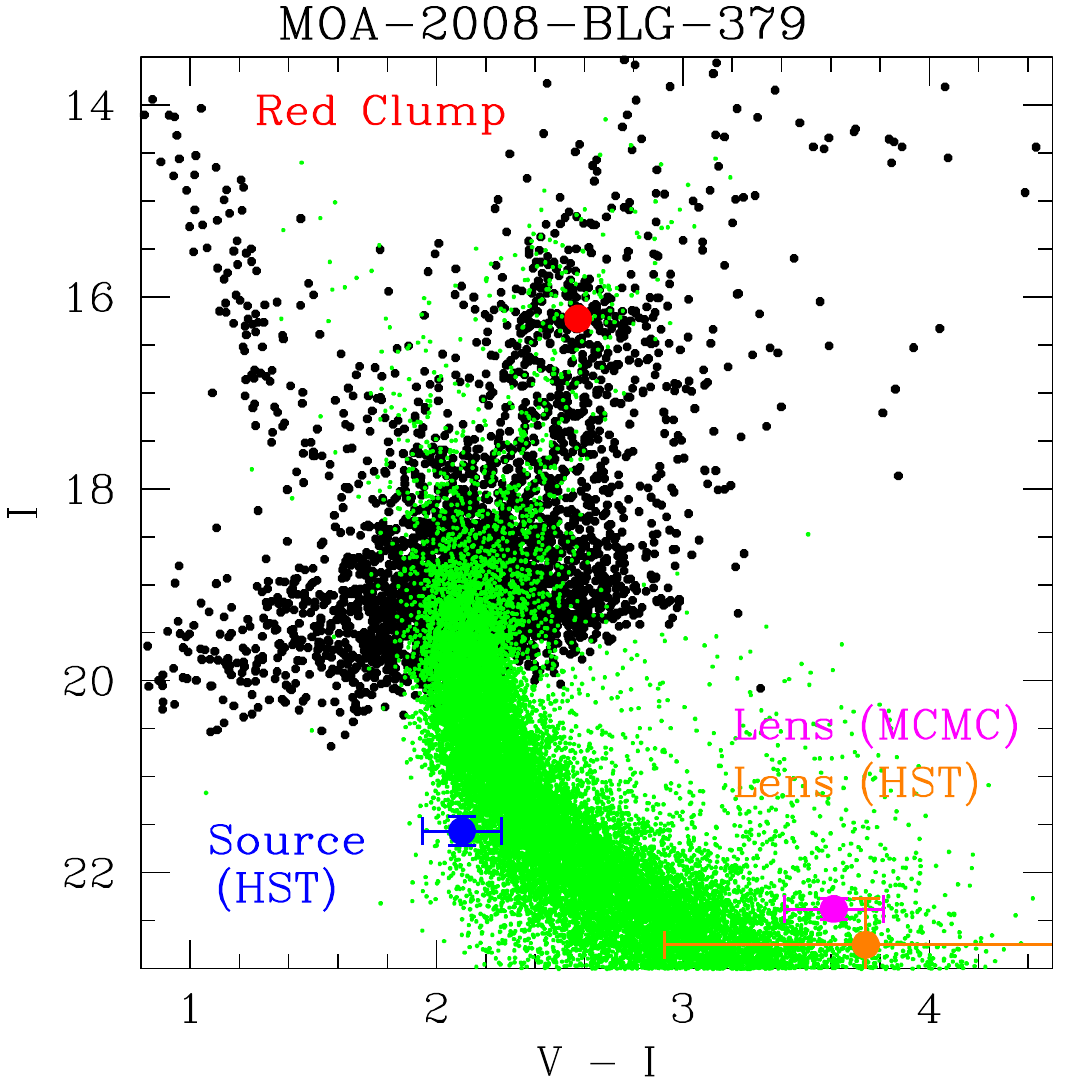}
\caption{The CMD of OGLE-3 stars within 90 arc seconds of microlensing event MOA-2008-BLG-379  (black dots),
with the {\sl Hubble} CMD of Baade's Window (green dots) \citep{holtzman98}, transformed to the same extinction and Galactic
bar distance as the MOA-2008-BLG-379 field. The
red spot is the red clump giant centroid; the source and lens magnitudes from our {\sl Hubble} observations
are indicated in blue and orange. As Table~\ref{tab-hst} indicates, the $V$-band (F555W) {\sl Hubble} images
detect the lens star (star 1) at $\sim 1.2\sigma$ significance, so the {\sl Hubble} measurements 
can be considered only an upper limit on the $V$-band brightness of the lens (and planetary host) star.
The magenta spot indicates the lens star color and magnitude inferred from our MCMC calculations
using the constrained \texttt{eesunhong} light curve modeling code. Since the lens star is likely
to be in the disk or the near side of the bulge, it is typically brighter than the bulge main sequence.
\label{fig-cmd}}
\end{figure}

\subsection{Interstellar Extinction}
\label{sec-extinct}

In order to apply the constraints from the high angular resolution follow-up images to the properties of
the star plus planet lens system, we must account for the extinction in the foreground of the lens, and 
we also need the extinction to the source star in order to determine the angular source size.
We determine the extinction in the foreground of the red clump giant stars following \cite{bennett14} 
using the red clump stars within $90^{\prime\prime}$ of the MOA-2008-BLG-379 event from
the OGLE-III photometry catalog \citep{ogle3-phot}. We identify the peak of the red clump stars color
magnitude distribution to be at $I_{\rm rcg} = 16.225 \pm 0.050$ and $(V-I)_{\rm rcg} = 2.575 \pm 0.030$,
as shown in Figure~\ref{fig-cmd}. Following \cite{nataf13}, we take the extinction corrected red clump giant
magnitude and color to be $I_{\rm rcg0} = 14.425$ and $(V-I)_{\rm rcg0} = 1.06$. This gives extinction
values of $A_I = 1.800$ and $A_V = 3.315$, implying a color excess of $E(V-I) = 1.515$. 
These values are within $0.5\sigma$ of the values quoted by \citet{suzuki14,suzuki14e}. We determine the
$K$-band extinction, $A_K = 0.182$, from the \cite{surot20} value of the color excess at the location
of MOA-2008-BLG-379, $E(J-K) = 0.369 \pm 0.0210$, using the \cite{nish06} infrared extinction law,
which gives $A_K/E(J-K)= 0.494\pm 0.006$. We assume that the extinction for the source star
is the same as the extinction of the center of the red clump giant distribution.

For the mass-luminosity relations, we must also consider the foreground extinction.
At a Galactic latitude of $ b = -3.1130^\circ$, and a lens distance of $\sim 4\,$kpc, the lens system
is likely to be behind most, but not all, of the dust that is in the foreground of the source. 
We assume a dust scale height of $h_{\rm dust} = 0.10\pm 0.02\,$kpc \citep{drimmel}, so that the
extinction in the foreground of the lens is given by
\begin{equation}
A_{i,L} = {1-e^{-|D_L(\sin b)/h_{\rm dust}|}\over 1-e^{-|D_S (\sin b)/h_{\rm dust}|}} A_{i,S} \ ,
\label{eq-A_L}
\end{equation}
where the index $i$ refers to the passband: $I$, $V$, or $K$. 

\section{Determination of Lens System Properties from Light Curve and High Angular Resolution Follow-up Data}
\label{sec-LC_hi_ang}

For MOA-2008-BLG-379, like a number of other planetary events, we find it useful to apply constraints from the
high angular resolution follow-up observations to the light curve models. This can prevent the light curve modeling 
from exploring parts of parameter space that are excluded by the high angular resolution follow-up observations.
There are multiple ways to use light curve modeling and high angular resolution follow-up observations to determine
the masses and distance of a planetary microlensing system. But, these methods can sometimes be compromised
by astrophysical complications or systematic measurement errors. So, it is generally useful to confirm mass and
distance measurements with multiple methods. 

\subsection{Light Curve Model and High Angular Resolution Image Parameters}
\label{sec-mod_im_param}

This section discusses the parameters that are important for determining the physical properties of 
planetary microlens systems from both light curve modeling and high angular resolution imaging. The
measurement of these parameters allows what is generally considered to be a ``full solution" for
a planetary microlensing system. The physical parameters that result from these ``full solutions"
include the masses of the lens masses (both stars and planets), their projected separation on the 
plane of the sky in physical units. In rare occasions, it is possible to determine more detailed properties
of planetary microlensing systems, such as the orbital inclination and eccentricity \citep{gaudi-ogle109,bennett-ogle109},
but this is much less likely for the cool, low-mass planets that microlensing is uniquely
sensitive to \citep{bennett96,bennett02}.
The parameters that are important for obtaining ``full solutions" for planetary microlensing events
are listed below, including parameters determined from both light curve modeling and 
high angular resolution imaging.
\begin{enumerate}
  \item Light curve model parameters:
  \begin{enumerate}
    \item The planet-star mass ratio, $q$. This is almost always measured with reasonable accuracy, but there
    are occasionally degeneracies, in which the light curve can be well fit by models with very different $q$ values.
    Some of these degeneracies can be resolved with high angular resolution follow-up imaging \citep{terry22}.
     \item The Einstein radius crossing time, $t_{\rm E}$. This is the time it takes for the lens-source relative motion
     to traverse the angular Einstein radius, $\theta_{\rm E}$. This is typically well measured, but there can be large
     uncertainties for faint source stars with planetary signals observed at high magnification
     because  $t_{\rm E}$ must be measured from the low-magnification part of the light curve  
     \citep{alard97,distefano95}. The microlensing parallax signal, discussed below in item 1(e), also tends to
     be found in the lower magnification parts of the light curve, so constraints on the microlensing parallax signal
     may also constrain $t_{\rm E}$.
     \item The source star magnitude and color, corrected for extinction, i.e.\ $I_{S0}$ and $(V_{S0}-I_{S0})$, 
     can be used to determine the source star's angular diameter, $\theta_*$ \citep{kervella_dwarf,boyajian14,adams18}.
     When the blending degeneracy \citep{alard97,distefano95} makes the $t_{\rm E}$ value very uncertain, 
     the source magnitude also has a large uncertainty, so the inferred source star angular diameter, $\theta_*$,
     inherits a large uncertainty.
     \item The source star radius crossing time, $t_*$ is a measure of finite source effects in a microlensing
     light curve. More than half of the known planetary microlensing events allow $t_*$ to be measured, and this allows the
     angular Einstein radius, $\theta_{\rm E} = t_{\rm E}\theta_*/t_*$, and the lens-source relative proper motion, 
     $\mu_{\rm rel,G} = \theta_*/t_*$, to be measured for most planetary events. 
     Most microlensing modeling uses the instantaneously geocentric inertial reference frame that moves with
     the Earth's velocity at the time of the event peak. We use the subscript, G, to indicate that this geocentric frame 
     has been used to measure the relative proper motion,  $\mu_{\rm rel,G}$.
     \item The microlensing parallax, $\bm{\pi}_{\rm E}$, is a two dimensional vector caused by the fact that the 
     microlensing event looks
     different from observers with different positions or velocities. This is most commonly observed due to the orbital motion
     of the Earth \citep{gould-par1,macho-par1}, but in some cases it can be measured by a satellite far from the 
     Earth \citep{udalski_ogle124} or from different observatories on the Earth \citep{gould09}. When the orbital
     motion of the Earth enables a  $\bm{\pi}_{\rm E}$ measurement for a microlensing event towards the
     Galactic bulge, the East component of $\bm{\pi}_{\rm E}$ is usually measured much more accurately
     than the North component, because the orbital acceleration of the Earth perpendicular to the line of sight 
     to the bulge is primarily in the East-West direction. This was the case situation for two previous planetary
     microlensing events with masses and distances determined with the help of high angular resolution
     follow-up observations \citep{aparna18,bennett_ogle71}.
   \end{enumerate}
   \item Parameters from high angular resolution imaging:
     \begin{enumerate}
     \item Excess flux at the location of the source star could be due to the lens star, but in some cases
     this can be due to a binary companion to the source or the lens, or even an unrelated star \citep{aparna17}.
     See \citet{koshimoto_bayes_follow} for a Bayesian method to address these issues.
     \item Lens star magnitude(s). 
     When the lens star has a measurable separation from the source star, it is possible to measure its brightness 
     with a much lower probability of confusion with a star other than the lens. Magnitude measurements in multiple passbands
     can provide a means for independent mass measurements that can be compared for consistency \citep{bennett15,batista15}.
     \item Source star magnitudes or magnitude limits. While the source star magnitudes are usually determined by the 
     microlensing light curve modeling, the blending degeneracy can interfere with the source magnitude determination,
     as mentioned in item 1(c), above. In these cases, source magnitude measurements or limits from high resolution imaging
     can be useful.
     \item The lens-source relative proper motion in the heliocentric coordinate system, $\mubold_{\rm rel,H}$. This is 
     determined from high angular resolution follow-up images when the lens-source separation can be measured. It can be
     used to help determine the microlensing parallax vector, $\bm{\pi}_{\rm E}$, because 
     $\bm{\pi}_{\rm E}\parallel \mubold_{\rm rel,G}$, but this requires a change from the heliocentric to geocentric
     coordinate systems.
     \end{enumerate}
\end{enumerate}

\subsection{Microlensing Event Mass-Distance Relations}
\label{sec-mass_dist}

Both the angular Einstein radius, $\theta_{\rm E}$, and the length of the microlensing parallax vector, $\pi_{\rm E}$, 
give relations that link the lens system mass to the lens and source distances, $D_S$ and $D_L$. These relations
are \citep{bennett_rev,gaudi_araa}:
\begin{equation}
M_L = {c^2\over 4G} \theta_{\rm E}^2 {D_S D_L\over D_S - D_L}  \ ,
\label{eq-m_thetaE}
\end{equation}
and 
\begin{equation}
M_L  =  {c^2\over 4G}{ {\rm AU}\over{\pi_{\rm E}}^2}{D_S - D_L\over D_S  D_L}  \ .
\label{eq-m_piE}
\end{equation}
Equations \ref{eq-m_thetaE} and \ref{eq-m_piE} can be combined to yield the lens mass in an expression 
with no dependence on the lens or source distance,
\begin{equation}
M_L = {c^2 \theta_{\rm E} {\rm AU}\over 4G \pi_{\rm E}} = {\theta_{\rm E} \over (8.1439\,{\rm mas})\pi_{\rm E}} \msun \ .
\label{eq-m}
\end{equation}
The lens system distance can also be determined from
\begin{equation}
D_L={{\rm AU}\over \pi_{\rm E}\theta_{\rm E}+1/D_S} \ ,
\label{eq-Dl}
\end{equation}
but it does depend on $D_S$. With clear measurements of both $\theta_{\rm E}$ and $\pi_{\rm E}$, it is possible 
to get a complete solution to a planetary microlensing event without the benefit of high angular resolution
imaging, but this is relatively rare. Strong $\pi_{\rm E}$ measurements are generally obtained only for 
relatively long duration events with bright source stars that occur toward the beginning or end of the 
Galactic bulge observing season, when the orbital acceleration of the Earth is approximately perpendicular
to the line of sight to the bulge \citep{muraki11}.

High angular resolution follow-up images images can allow the source and lens stars to be resolved \citep{bennett15,batista15,van20}
or partially resolved \citep{aparna18,bennett_ogle71} and enable their
magnitudes to be measured. A measured magnitude of the lens star yields the following relation when the $K$-band
brightness of the lens is measured
\begin{equation}
K_L = 10 + 5\log_{10} (D_L/1\, {\rm kpc}) + {\cal M}_K(M_L) + A_{K,L} \ ,
\label{eq-nass_lum}
\end{equation}
where ${\cal M}_K(M_L)$ is a $K$-band mass-luminosity relation. This requires the knowledge of the dust extinction, $A_{K,L}$  in
the foreground of the lens star. In most cases, an empirical mass-luminosity relation for a main sequence star is
appropriate, but for host stars of $\sim 1\msun$, the luminosity may change significantly over the age of
the Galaxy, so a collection of isochrones is likely to be more accurate \citep{beaulieu16,van20}. 
Mass-luminosity relations in multiple passbands can be used to confirm the mass measurement \citep{bennett15,batista15},
but they can also be used to identify circumbinary planets \citep{bennett16}, since the binary star systems have
redder colors than single stars of the same mass \citep{terry21}.

These same high angular resolution follow-up images that resolve or partially resolve the lens and source stars
can also be used to confirm the identification of the lens star by measuring the lens-source relative proper motion,
$\mubold_{\rm rel}$, which can be compared to the magnitude of the relative proper motion vector, $\mu_{\rm rel,G} = \theta_*/t_*$,
which can often be determined from the angular source star radius, $\theta_*$, and source radius crossing 
time, $t_*$, from the light curve model. However, these two independent $\mubold_{\rm rel}$ values are not measured
in the same reference frame. The light curve model provides $\mu_{\rm rel,G}$ in the instantaneously geocentric inertial 
reference frame that moves with the earth at the time of peak magnification, while the high angular resolution follow-up
imaging gives the 2-dimensional vector proper motion, $\mubold_{\rm rel,H}$, in the heliocentric reference frame 
(plus a small correction due to geometric parallax, which is usually negligible). The 2-dimensional vector proper motions
in the different reference frames
are usually quite similar, but the difference can be significant if the relative proper motion or the lens distance, $D_L$, is
small. The geocentric relative proper motion, $\mubold_{\rm rel,G}$ can be determined with the following formula
\citep{dong-ogle71}:
\begin{equation}
\mubold_{\rm rel,G} = \mubold_{\rm rel,H} - \frac{{\bm v}_{\oplus} \pi_{\rm rel}}{\rm AU}  \ ,
\label{eq-mu_helio}
\end{equation}
where ${\bm v}_{\oplus}$ is the projected velocity of the earth relative to the sun (perpendicular to the 
line-of-sight) at the time of peak magnification. The projected velocity for MOA-2008-BLG-379 is
${{\bm v}_{\oplus}}_{\rm E, N}$ = (19.680, -2.5983) km/sec = (4.152, -0.548) AU/yr at the peak of the 
microlensing light curve, HJD'= 4688. 
The relative parallax is 
defined as $\pi_{\rm rel} \equiv 1/D_L - 1/D_S$, where $D_L$ and $D_S$ are lens and source 
distances, so equation~\ref{eq-mu_helio} can be written as:
\begin{equation} 
\mubold_{\rm rel,G} = \mubold_{\rm rel,H} - (4.152, -0.548)\times (1/D_L - 1/D_S) \ ,
\label{eq-mu_helio2}
\end{equation}
when $D_L$ and $D_S$ are given in units of kpc, and $\mubold_{\rm rel,H}$ and $ \mubold_{\rm rel,G}$
are in units of mas/yr. So, a precise comparison of $\mubold_{\rm rel,H}$ from high angular resolution follow-up
observations to $\mu_{\rm rel,G}$ from the light curve model requires some knowledge of $D_L$ and $D_S$, but
in many cases, a precise comparison may not be needed. For example, \citet{aparna17} found that a candidate
host star for the planet MOA-2008-BLG-310Lb was moving toward the source instead of away from it, after the
event. This showed that the likely host star suggested by \citet{janczak10} was actually not related to the 
microlensing event. This same argument was used to exclude a main sequence candidate for the 
MOA-2010-BLG-477L host star \citep{blackman_477}, leading to the conclusion that this lens system
is the first example of a planet in a wide orbit about a white dwarf host star.

The measurement of $\mubold_{\rm rel,H}$ is also very useful for the determination of precise values for the
microlensing parallax parameter, ${\bm{\pi}_{\rm E}}$. In most cases, one component of this 2-dimensional
vector is measured more precisely than the other. This is the component  of ${\bm{\pi}_{\rm E}}$ that is 
parallel to the orbital acceleration
of the observer, and for microlensing events observed towards the Galactic bulge, the direction that is measured more precisely
is quite close to the East-West direction, so it is usually the case that the East component
of ${\bm{\pi}_{\rm E}}$ is measured precisely, while the North component is only weakly constrained.
Fortunately,
the microlensing parallax vector, ${\bm{\pi}_{\rm E}}$ is parallel to the $\mubold_{\rm rel,G}$ vector, 
which can often be determined very precisely, using eq.~\ref{eq-mu_helio}, when we have a good
measurement of $\mubold_{\rm rel,H}$.
The microlensing parallax and Geocentric relative proper motion are related by
\begin{equation}
{\bm{\pi}_{\rm E}} = \frac{\pi_{\rm rel}}{t_{\rm E}}\frac{\mubold_{\rm rel,G}}{|\mu_{\rm rel,G}|^2} \ ,
\label{eq-piE_muG}
\end{equation}
so with measurements of $\pi_{\rm E,E}$ and $\mubold_{\rm rel,H}$, we can use equations~\ref{eq-mu_helio}
and \ref{eq-piE_muG} to solve for $\pi_{\rm E,N}$ \citep{gould-diskorhalo,ghosh04,bennett07}. 
This leads to a quadratic equation in order to solve for  $\pi_{\rm E,N}$ \citep{gould-1dpar}, but in general, there is no ambiguity
between the two solutions, as one solution either requires a negative lens distance, $D_L$, or predicts 
a lens brightness that is strongly inconsistent with the measured lens magnitude \citep{aparna18}.
This method was used to solve for $\pi_{\rm E,N}$ to yield a precise measurement of the ${\bm{\pi}_{\rm E}}$ vector
for both OGLE-2005-BLG-071 \citep{bennett_ogle71} and OGLE-2012-BLG-0950 \citep{aparna18}, and in 
both cases, the microlensing parallax measurement confirmed the lens system masses and distance
indicated by the host star brightness and $\theta_{\rm E}$ values.

\subsection{Applying Constraints from High Angular Resolution Follow-up Observation on Light Curve Models}
\label{sec-LC}

In principle, one can determine the physical parameter of the host star plus planet lens system with 
independent analyses of the light curve and high angular resolution follow-up observations. This has been 
done for the planetary microlensing event OGLE-2005-BLG-169 \citep{bennett15,batista15}, but there
are several potential problems with this approach. First, it can be the case that the follow-up data
restrict the parameters of the lens system to a very small fraction of the parameter space volume that 
was consistent with the observed light curve. This then makes Markov Chain Monte Carlo (MCMC) calculations
of the distribution of light curve parameters very inefficient since most of the light curve models accepted
by the Markov Chain are excluded by the follow-up observation constraints. This is particularly true
for events that have partial measurements of the microlensing parallax effect, due to the orbital motion
of the Earth \citep{aparna18,bennett_ogle71}. Since the microlensing parallax vector points in the
same direction as the lens-source relative proper motion vector, the follow-up observations exclude
a large fraction of the models that are consistent with the light curve data.

There are also a variety of both subtle microlensing features and systematic photometry errors that are easier to diagnose 
with the help of the high angular resolution imaging data. Microlensing parallax is one such feature that is present
in every light curve produced by a telescope in a heliocentric orbit, but the microlensing parallax signal is often too 
weak to be clearly detected. If the source star is in a binary system, then it can have orbital motion that also affects the 
light curve, similar to the microlensing parallax due to the orbital motion of the observer. 
This is known as xallarap. A binary companion to the source can also be microlensed, but this
possibility is usually not considered, unless the companion has a dramatic influence on the light curve
\citep{bennett-moa117} or if there is some danger of a binary source feature being interpreted as a planetary signal
\citep{gaudi97,ogle390}. The orbital motion of the planet can also have a significant effect on the light curve,
but there is often some degeneracy between the orbital motion and the microlensing parallax 
\citep{gaudi-ogle109,bennett-ogle109} or xallarap parameters. Also, all three of these features (microlensing
parallax, xallarap, the lensing of a binary companion to the main source star) can be mimicked by low-level
systematic photometry errors.

Another problem can occur for high magnification events with faint source stars, like the one analyzed in this paper, 
MOA-2008-BLG-379. High magnification events are extremely sensitive to planetary signals \citep{griest98,rhie00},
and the faintness of the source star makes it easier to detect the lens stars, which are usually fainter than the 
source stars. However, it can be a challenge to determine the brightness of the source stars for such events, because
of a degeneracy between Einstein radius crossing time ($t_{\rm E}$) and source brightness of microlensing events
\citep{alard97,distefano95}, which can only be resolved with relatively high precision photometry obtained
at low magnification. Thus, the measurement of $t_{\rm E}$ and the source brightness is sensitive to low-level
systematic photometry errors. 
Furthermore, the shape of the light curves at low magnification can also depend on 
microlensing parallax effects, so it is prudent to include microlensing parallax in the light curve modeling, because
the orbital motion of the Earth always produces a microlensing parallax signal that could affect the light curve
constraints on $t_{\rm E}$ and the source brightness.

In order to address this problem, we have modified our fitting code \citep{bennett96,bennett-himag}, which now
goes by the name, \texttt{eesunhong}\footnote{\url{https://github.com/golmschenk/eesunhong}}, 
in honor of the original co-author of the code \citep{rhie_phystoday,rhie_obit}.
This new version of \texttt{eesunhong} includes the constraints on the brightness and separation of the lens and source stars
from the high angular resolution follow-up images from {\sl Keck} AO and {\sl Hubble}. However, in order to determine 
the mass of the host star based on the lens-source relative proper motion, which determines the angular 
Einstein radius, $\theta_{\rm E}$, we need to know the distance to the source star, $D_S$, so that we can
use the mass-distance relation given in Equation~\ref{eq-m_thetaE}.
This requires us to include the source distance, $D_S$, as a light curve model parameter, and we include a
weighting from the \citet{koshimoto_gal_mod} Galactic model as a prior for the $D_S$ parameter. 
We also use the \citet{koshimoto_gal_mod} Galactic model  to provide a prior for the distance to the lens
for a given value of the $D_S$ parameter. However, this prior for $D_S$ at fixed $D_L$ is used to weight 
the entries in a sum of Markov chain values, rather than directly in the light curve modeling code.

The light curve modeling code does use constraints 
from the {\sl Keck} analysis for $\mubold_{\rm rel,H}$ that are given in Table~\ref{tab-Keck} 
and  on the lens magnitude, $K_{L} = 18.815 \pm 0.106$, based on
with our identification from the {\sl Hubble} analysis that star 1 is the lens 
(and planetary host) star. The $K_L$ error bar is larger than the value of 0.056 quoted for the $K_L$
measurement in Section~\ref{sec-Keck} because we have added a $K$-band mass-luminosity relation uncertainty
of 0.09 mag in quadrature to the measurement uncertainty.
This constraint is implemented with a Gaussian distribution $\chi^2$ contribution to the
total model $\chi^2$. The measured {\sl Hubble} source and lens magnitudes are 
$I_S = 21.557 \pm 0.149$ and $I_L = 22.742 \pm 0.488 \pm 0.190$, where the $\pm 0.190$ uncertainty
is our estimate of the $I$-band mass-luminosity relation uncertainty. This implies $I_L =  22.742 \pm 0.524$.
We also apply a constraint to the
combined brightness of the lens and source, as this is measured more precisely than the individual lens and
source magnitudes. Our {\sl Hubble} measurement finds  $I_{S+L} = 21.243\pm 0.011$, but we add a systematic
uncertainty of 0.10 mag to this value to account for the mass-luminosity relation uncertainty for the lens star and 
any systematic error that might be caused by measurement of the combined brightness of two partially resolved stars.
This yields our constraint value of $I_{S+L} = 21.243\pm 0.101$.
The light curve does not provide a good measurement of the
source $V$-band magnitude, so we do not attempt to constrain that, but the {\it Hubble} data do provide an 
upper limit on the $V$-band brightness of the lens star, which is a lower limit on the magnitude: 
$V_L \geq 26.493 \pm 0.684$. This limit implies
a Gaussian contribution to $\chi^2$ for models with $V_L < 26.493$ with no $\chi^2$ contribution
for models with $V_L \geq 26.493$.


\begin{deluxetable}{cccccc}
\tablecaption{Best Fit Model Parameters with $\mubold_{\rm rel,H}$ and Magnitude Constraints
                         \label{tab-Cmparams} }
\tablewidth{0pt}
\tablehead{
& \multicolumn{2}{c} {$u_0 < 0$} & \multicolumn{2}{c} {$u_0 > 0$} &  \\
\colhead{parameter}  & \colhead{$s<1$} & \colhead{$s> 1$} & \colhead{$s<1$} & \colhead{$s> 1$} &\colhead{MCMC averages}
}  
\startdata
$t_{\rm E}$ (days) & 55.637 & 55.087 & 55.452 & 55.527 & $55.8\pm 5.5$  \\   
$t_0$ (${\rm HJD}^\prime$) & 4687.8953 & 4687.8742 & 4687.8952 & 4687.8739 & $4687.8795\pm 0.0091$  \\
$u_0$ & $-0.0047088$ & $-0.0055301$ & 0.0047275 & 0.0055000 & $-0.00529\pm 0.00054$  \\
  & & &  \multicolumn{2}{c} {($u_0 > 0$)} & $0.00531\pm 0.00054$  \\
$s$ & 0.92651 & 1.08670 & 0.92706 & 1.08847 & $0.929 \pm 0.007$  \\
  & & &  \multicolumn{2}{c} {($s > 1$)} &  $1.086\pm 0.007$  \\
$\alpha$ (rad) & -1.13442 & -1.13131 & 1.13417 & 1.13111 & $-1.1320\pm 0.0023$  \\
  & & &  \multicolumn{2}{c} {($u_0 > 0$)} &  $1.1319\pm 0.0023$  \\
$q \times 10^{3}$ & 5.2320 & 5.3139 & 5.2398 & 5.2687 & $5.37 \pm 0.41$  \\
$t_\ast$ (days) & 0.02191 & 0.02216 & 0.02183 & 0.02219 & $0.0221\pm 0.0008$ \\
$\pi_{\rm E,N}(t_{\rm fix} = 4688)$ & 0.07313 & 0.07442 & 0.07518 & 0.07541 & $0.080\pm 0.026$ \\
$\pi_{\rm E,E}(t_{\rm fix} = 4688)$ & 0.19663 & 0.19697 & 0.19594 & 0.19540 & $0.207\pm 0.037$\\
$D_{s}$ (kpc) & 8.2542 & 8.0907 & 8.2568 & 8.3499 & $8.11\pm 1.32$ \\
fit $\chi^2$ & 1291.94 & 1290.01 & 1292.24 & 1290.07 &  \\
dof & $\sim 1276$ & $\sim 1276$ & $\sim 1276$ & $\sim 1276$ \\
\enddata
\end{deluxetable}

Table~\ref{tab-Cmparams} shows the parameters of our four degenerate light curve models and
the Markov Chain average of all four models. The parameters that apply to single lens models are
the Einstein radius crossing time, $t_{\rm E}$, the time of closest alignment between the source and the
lens system center-of-mass, $t_0$, and the distance of closest approach between the source and the lens 
system center-of-mass, $u_0$, which is given in units of the Einstein radius. The addition of a second lens mass
requires three additional parameters, the mass ratio of the two lens masses, $q$, their separation, $s$, in
units of the Einstein radius, and angle, $\alpha$, between the source trajectory and the transverse 
line that passes through the two lens masses. In addition, a large fraction of binary lens systems exhibit
finite source effects that can be modeled with the addition of the source radius crossing time parameter, $t_*$.
We include the North and East components of the microlensing parallax vector $\pi_{\rm E,N}$ and $\pi_{\rm E,E}$
that are defined in an inertial ``geocentric" coordinate system that is fixed to the 
Earth's orbital velocity at $t_{\rm fix}  = 4688$. For each passband (MOA-red, OGLE-$I$ and OGLE-$V$) there
are two linear parameters to describe the source flux and the blend flux (which accounts for blended starlight
that is not absorbed in the $\sim$uniform sky background. Following \citet{rhie_98smc1}, the source and
blend fluxes are determined by a linear fit to the model with all the other parameters fixed.
These constrained models have 3617 observations, 
10 non-linear parameters, 6 linear parameters and 8 constraints for a total of 3609 degrees of freedom.

For high magnification events, like MOA-2008-BLG-379, the transformation $s \rightarrow 1/s$ often 
has only a slight change on the shape of the light curve. This is often referred to as close-wide degeneracy \citep{dominik99},
and it applies to MOA-2008-BLG-379. However, as with many other events, the MOA-2008-BLG-379 does not
strictly meet the close-wide degeneracy conditions that make the central caustic nearly identical under the
$s \leftrightarrow 1/s$ transformations. Of course, a microlensing light curve only samples a fraction of the
microlensing magnification pattern, so this is not really a surprise. \cite{zhang22} have examined this situation
systematically, and have explained in more detail the conditions needed for this degeneracy, which they refer
to as the offset degeneracy (although the term ``central caustic offset degeneracy" would be more descriptive).

This event, like most Galactic bulge microlensing events is also subject to the ecliptic degeneracy
\citep{poindexter05}, which is exact for events in the ecliptic plane. This degeneracy involves replacing
a binary lens system with its mirror image, and it is the orbital motion of the Earth, which is detected via the 
microlensing parallax effect, that breaks the mirror symmetry. The models with the different lens system orientations
have opposite signs for the $u_0$ and $\alpha$ parameters.
The light curve data for MOA-2008-BLG-379
does not provide a strong signal for the microlensing parallax effect, and we have only included microlensing
parallax in our modeling because the high angular resolution imaging constrains the microlensing parallax
parameters and these parameters might be correlated with other model parameters. The best fit models
that differ by this ecliptic degeneracy (with $u_0 < 0$ and $u_0 > 0$) are nearly identical, but the best fit
wide model with a planet-star projected separation of $s = 1.08132$ is a slightly better fit than the
best fit close model with $s = 0.93472$ by $\Delta\chi^2 = 1.93$.

In order to check the consistency of the high angular resolution observation constraints with the 
light curve data, we can compare the $\chi^2$ values for the best fit models with and without these constraints.
The best fit constrained model has $\chi^2 = 1290.01$ for 1287 light curve photometry measurements, while the
best unconstrained model has $\chi^2 = 1285.27$, for a difference of $\Delta\chi^2 = 4.74$. A total of eight
constraints were imposed on the light curve models. These were
constraints on two components of $\mubold_{\rm rel,H}$, three constraints on the lens star magnitudes
($K_L$, $I_L$, and $V_L$), one constraint of the source star magnitude, $I_S$, one constraint on the 
combined source plus lens star magnitude, $I_{S+L}$, and one constraint on the source star distance, $D_S$.
This $\Delta\chi^2 = 4.74$ increase had contributions of 1.59 from the light curve fit, 0.41 from the $\mubold_{\rm rel,H}$
constraint, 1.87 from the 3 lens magnitude constraints, 0.08 from the $I_S$ constraint, 0.70 from the 
$I_{S+L}$ constraint, and 0.10 from the source distance constraint. Thus, there appear to be no conflict between
the light curve data used in the analysis and the constraints from the high angular resolution follow-up observations.

\section{Lens Properties}
\label{sec-lens_prop}

 \begin{deluxetable}{cccc}
\tablecaption{Measurement of Planetary System Parameters from the Lens Flux Constraints\label{tab-params}}
\tablewidth{0pt}
\tablehead{\colhead{parameter}&\colhead{units}&\colhead{values \& RMS}&\colhead{2-$\sigma$ range}}
\startdata
Angular Einstein Radius, $\theta_{\rm E}$&mas&$0.754\pm 0.040 $&0.672--0.832 \\
Geocentric lens-source relative proper motion, $\mu_{\rm rel, G}$&mas/yr&$5.042\pm  0.149$&4.70--5.29\\
Host star mass, $M_{\rm host}$&${\msun}$&$0.434\pm 0.065$ & 0.307--0.561\\
Planet mass, $m_p$&$M_{\rm Jup}$& $2.44\pm 0.49$ & 1.56--3.47\\
Host star - Planet 2D separation, $a_{\perp}$&AU&$2.70\pm 0.42$ & 1.87--3.53\\
Host star - Planet 3D separation, $a_{3\rm d}$&AU&$3.3^{+1.8}_{-0.6}$& 2.1--12.9\\
Lens distance, $D_L$&kpc &$3.44\pm 0.53$& 2.44--4.53\\
Source distance, $D_S$ &kpc & $7.77\pm 1.27$& 5.19--10.24\\
\enddata
\end{deluxetable}

\begin{figure}
\epsscale{1.0}
\plotone{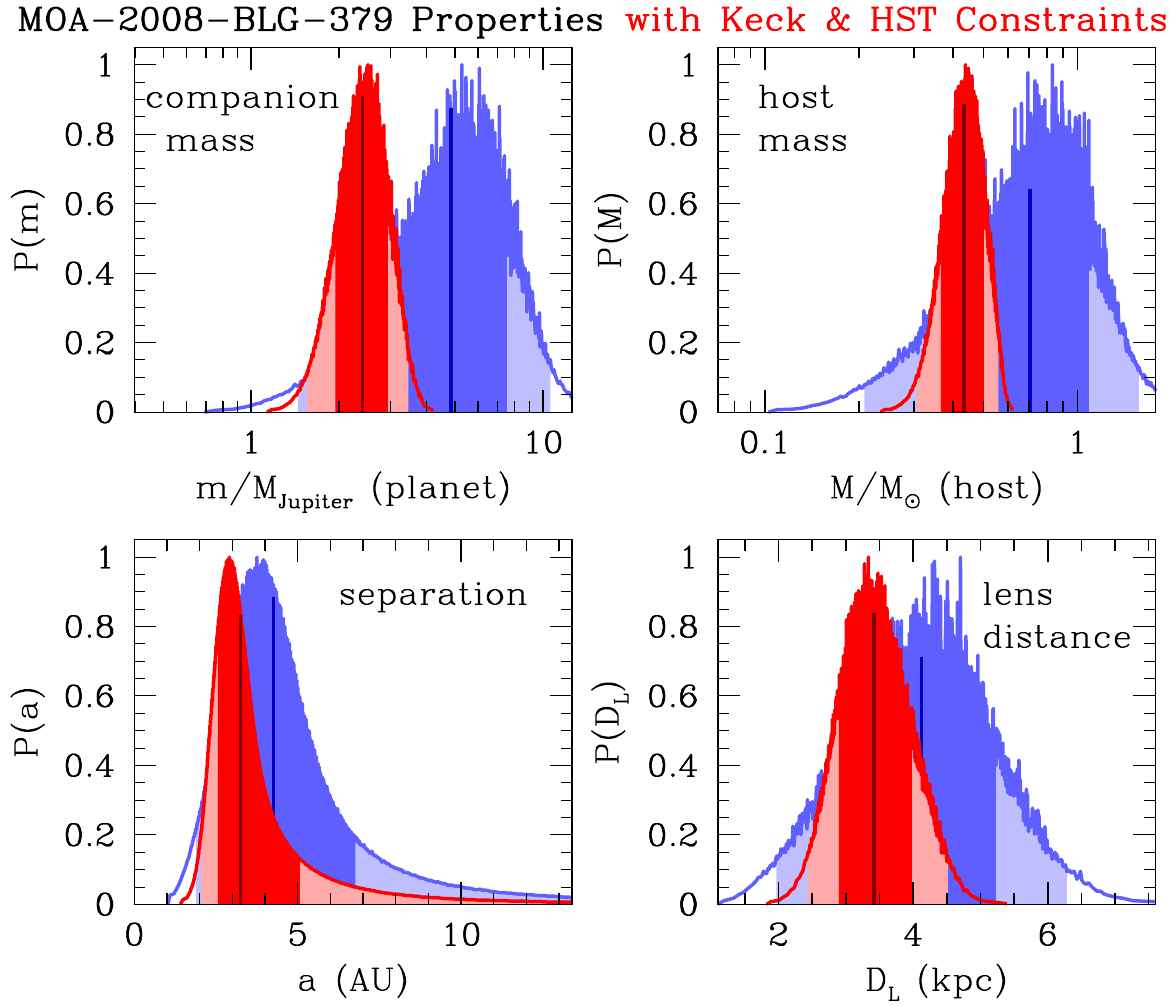}
\caption{The Bayesian posterior probability distributions for the planetary companion mass, host mass, 
their separation and the distance to the lens system are shown with only light curve constraints in blue 
and with the additional constraints from our {\sl Keck} and {\sl Hubble} follow-up observations in red.
The central 68.3$\%$ of the distributions are shaded in darker colors (dark red and dark blue) and the 
remaining central 95.4$\%$ of the distributions are shaded in lighter colors. The vertical black line marks 
the median of the probability distribution of the respective parameters.}
\label{fig-lens_prop}
\end{figure}

 Table~\ref{tab-params} and Figure~\ref{fig-lens_prop} provide the results of our analysis. These results
were obtained by
summing over the MCMC results that are summarized in Table~\ref{tab-Cmparams} to determine the 
posterior distribution of the properties of the MOA-2008-BLG-379L planetary system. We have run four
Markov chains for each of the $\chi^2$ minima listed in Table~\ref{tab-Cmparams}, and we have
applied a weight of $e^{-\Delta\chi^2/2}$ to the Markov chains, with $\Delta\chi^2$ defined as the
difference between the best fit $\chi^2$ for each $\chi^2$ minima compared to the overall best
fit $\chi^2$, which was the $u_0 < 0$, $s > 1$ model. With the burn-in phases of the Markov chains
removed, there were a total of 105,0161 accepted Markov chain steps used in these calculations.

Because we constrained lens-source relative proper motion, $\mubold_{\rm rel,H}$, the host star $K$, $I$, and $V$ 
magnitudes, and the combined host and source star magnitudes in the light curve modeling, we do not 
apply these constraints when summing the MCMC results. Also, since the source distance, $D_S$, prior was 
applied to the light curve model, we do not apply it again. However, we do use a Galactic model 
prior on the lens distance, $D_L$, for the $D_S$ value for each light curve model, constrained
by the measured $\mubold_{\rm rel,H}$ value.

Thus far, we have assumed that all stars are equally likely to host the planet with the measured mass ratio, $q$.
This is a common assumption, but we do not have any empirical evidence that it is true. In fact, as mentioned
in the introduction, the preliminary evidence from both microlensing and radial velocity surveys indicates that
the planet hosting probability scales in proportion to the host star mass. Therefore, we have applied a prior
proportional to $M_{\rm host}$ to our sum over the MCMC results. Fortunately, because the light curve
and high resolution imaging data constrain the mass, this prior has a small effect on the results. The 
results presented in Table~\ref{tab-params} and Figure~\ref{fig-lens_prop} change by $< 0.2\sigma$, if
we switch to the more common (but likely incorrect) prior assumption that the planet hosting probability
is independent of host mass.

We find that the
host star has a mass of $M_{\rm host} = 0.434\pm 0.065\msun$ and it is orbited by a super-Jupiter mass 
planet with $m_p = 2.44\pm 0.49 M_{\rm Jup}$ at a projected separation of $a_\perp = 2.70\pm 0.42\,$AU.
This translates to a three-dimensional separation of $a_{3\rm d} = 3.3^{+1.8}_{-0.6}\,$AU under the
assumption of a random orientation of the planetary orbit, and the lens
system is located at a distance of $D_L = 3.44\pm 0.53\,$kpc. These distributions are indicated by the 
red histograms in Figure~\ref{fig-lens_prop}.
These results are a dramatic improvement in precision over blue histograms that indicate the 
parameters predicted by our Bayesian analysis without any constraints from {\sl Keck} or {\sl Hubble} observations. 

While high angular resolution follow-up observations provide much more precise determinations of the 
properties of the planetary system responsible for the microlensing event, there is one significant inconsistency
between the analysis with and without the high angular resolution follow-up observations. We have found the
Einstein radius crossing time to be $t_{\rm E} = 55.8\pm 5.5\,$days, which is noticeably larger than the value
of $t_{\rm E} = 45.0\pm 6.2\,$days obtain in our analysis without the high angular resolution follow-up observation
constraints and the discovery paper \citep{suzuki14,suzuki14e} value of $t_{\rm E} = 42.3\pm 0.5\,$days. The
longer $t_{\rm E}$ value is due to the fact that the {\sl Hubble} and {\sl Keck} data imply a source that is 0.32 magnitudes fainter
than the value from the discovery paper. The discovery paper used
the unphysical constraint: ${\bm{\pi_{\rm E}}}\equiv 0$, which could, in some cases, lead to unphysically small error bars
on $t_{\rm E}$ \citep[e.g.][]{sumi10,batista11}, but in this case, the small error bars were reported because 
the Markov Chain Monte Carlo analysis was not run long enough to have converged.
This constraint does not come directly from the {\sl Hubble} $I_S$ measurement due to the large correlated uncertainty in the fractions
of $I$-band flux attributed to the source and lens stars. This is due to their relatively small separation at the time of the 
{\sl Hubble}  observations. The fainter source star is implied by the combination of constraints from all three 
passbands including the tight constraint on the combined $I$-band brightness of the lens and source stars,
as well as the lens-source relative proper motion,  $\mubold_{\rm rel,H}$, constraint. If the lens-source separation 
had been larger at the time of the {\sl Hubble} images, the direct $I_S$ measurement would be more precise. In this
case, it would provide a precise constraint on $t_{\rm E}$.

Despite the inconsistency between our models and the Einstein radius crossing time, $t_{\rm E}$, and 
the source brightness values from \cite{suzuki14,suzuki14e}, our results for the lens system masses and distance
fall within the ranges $M_{\rm host} = 0.56^{+0.24}_{-0.27}\msun$, $m_p = 4.1^{+1.7}_{-1.9} M_{\rm Jup}$,
and $D_L = 3.3^{+1.3}_{-1.2}\,$kpc quoted in that paper. This is partly because the uncertainties in these
parameters are large without the follow-up observations, but also because the larger $t_{\rm E}$
partially compensates for the smaller angular source star radius, $\theta_*$, in the calculation of the angular Einstein radius,
$\theta_{\rm E} = t_{\rm E}\theta_*/t_*$.

\section{Lessons for Modeling Roman Galactic Exoplanet Survey Events}
\label{sec-RGES_lessons}

The original version of this paper attributed a $\sim 30\sigma$ difference between the Einstein radius crossing time
predicted by the MOA-2008-BLG-379Lb discovery paper \cite{suzuki14,suzuki14e} to the fact that the models
in this paper fixed the microlensing parallax parameter to ${\bm{\pi_{\rm E}}}\equiv 0$. This is a common 
procedure, but it could lead to unphysical constraints on the other light curve parameters. However, as we explain
in Appendix~\ref{sec-old-lc}, this is not the case for the MOA-2008-BLG-379 event. However, this is the case
for three other events (out of 28) in the \citet{suzuki16} statistical sample. 
The $t_{\rm E}$ values for microlensing parallax models differ from the values with ${\bm{\pi_{\rm E}}}\equiv 0$
by $17\sigma$, $12\sigma$, and $6.3\sigma$ for OGLE-2007-BLG-368 \citep{sumi10}, MOA-2009-BLG-387 \citep{batista11},
and OGLE-2012-BLG-0563 \citep{fukui15}, respectively. This difficulty in determining $t_{\rm E}$ and the source star magnitude
for high magnification events with faint sources is a general one, due to the fact that it is the low-magnification parts a single lens
light curve provide the strongest constraints on $t_{\rm E}$ and the source brightness \citep{yee12}.
This may play a role in the failure of the ${\bm{\pi_{\rm E}}}\equiv 0$ modeling to produce a reasonable $t_{\rm E}$ error bar
for OGLE-2012-BLG-0563,
but there is likely to be a different source of the $t_{\rm E}$ error bar problems for OGLE-2007-BLG-368 and MOA-2009-BLG-387,
since these event have more modest magnification.
Both of these events have strong, caustic crossing features which strongly break the circular symmetry of a single lens
magnification pattern. This makes these events much more sensitive to microlensing parallax effects.

The analysis of the MOA-2008-BLG-379 high resolution imaging data
has illustrated some ways in which light curve modeling can be improved when one
is trying to determine the physical parameters of the lens system with the
help of high angular resolution imaging to measure the brightness and separations of the lens stars from
the microlensing source stars. 
The high angular resolution images obtained for the exoplanet microlensing
survey of the {\sl Roman Space Telescope} \citep{bennett18_wfirst,penny19,johnson20}, have long been thought
to play a key role in the measurement of host star and planet masses for the planetary systems 
discovered in {\sl Roman's} Galactic Bulge Time Domain Survey \citep{bennett02,bennett07}. Our
{\sl Keck} and {\sl Hubble} observations of the microlens planetary systems from the \citet{suzuki16} statistical
sample have found that the previous analyses of most of these planetary microlensing events had to be
redone in order to determine the masses and distance of these lens systems.

The inclusion of constraints from the high angular resolution imaging in the light curve modeling has proved to be very 
useful. For our analysis of MOA-2008-BLG-379, modeling with these constraints has helped to recognize the importance
of including microlensing parallax in the modeling even when there is little or no evidence of a measurable
microlensing parallax signal in the data. This is because constraining $\piEbold = 0$ can impose incorrect constraints
on $t_{\rm E}$ and the source brightness. However, it can also be necessary to avoid unphysically large $\piEbold$ values
by applying a prior based on a Galactic model.
Similar constraints have made the modeling of events with 
accurate measurements of only one component of the 2-dimensional $\piEbold$ vector much more efficient 
by excluding a very large fraction of models that are consistent with the light curve, but not consistent with the direction
of lens-source relative proper motion, $\mubold_{\rm rel,H}$, measured from the high angular resolution 
follow-up data \citep{aparna18,bennett_ogle71}.

Since the very first discovery of a microlensing event,
it has been quite common for modelers to ignore the effects of ``higher order effects" when they
do not seem necessary to explain the data. These higher order effects include
microlensing parallax, planetary orbital motion, additional lens mass, and a binary companion to the source, which may also
be microlensed or can generate detectable orbital motion for the source star, which is known
as ``xallarap\rlap". This procedure, favoring the minimal model that can explain the data, is considered preferable
when establishing the detection of some new light curve feature, such as the first microlensing parallax
detection \citep{macho-par1} or the first exoplanet found by microlensing \citep{bond04}. In fact, the
binary lens nature of the very first microlensing event detected \citep{dominik-lmc1,rhie-lmc1} was
ignored in the discovery paper \citep{macho-lmc1}. In the case of MOA-2008-BLG-379, the apparent 
systematic photometry errors leading an unphysically large $\pi_{\rm E}$ values provided an additional 
motivation to exclude microlensing parallax from the light curve models.

However, we know that every planetary microlensing event has both microlensing parallax and planetary
orbital motion. All telescopes are on orbits about the Sun and bound planets orbit their host stars,
so these effects will always be present. Even if the data do not have sufficient precision to measure these
effects, they can still influence the uncertainties on other parameters of interest, such as the source star
brightness and Einstein radius crossing time ($t_{\rm E}$), as we have seen for this event. The situation
is somewhat similar for planetary event MOA-2007-BLG-192, which involved a source star only slightly
brighter than the MOA-2008-BLG-379 source. This event was published with a large $\piEbold$  value \citep{bennett08}
that turned out to be contaminated by a systematic error \citep{koshimoto_blg_pl,terry_mb07192}
due to the color dependence of atmospheric 
refraction \citep{bennett12}. This error would likely have been found earlier if a prior distribution for 
$\piEbold$ had been used in the modeling.

The orbital motion of the planet can often produce effects similar to microlensing parallax
\citep{bennett-ogle109,sumi16}, so most of the planets in the \citet{suzuki16} sample with microlensing
parallax measurements, also have had planetary orbital motion included in their models
\citep{dong-ogle71,gaudi-ogle109,muraki11,batista11,bachelet12,moa328,skowron15,bennett16,bennett-moa117}.
So, the orbital motion of the planetary (and possible stellar companions) to the host star should generally
be included in the modeling to ensure an unbiased $\piEbold$  measurement.

Another astrophysical effect that can interfere with $\piEbold$  measurements is a binary companion to the 
source, which can induce source star orbital motion in a similar manner to the way that the Earth's orbital 
motion can reveal the microlensing parallax effect. However, this is orbital motion at the location of the 
source rather than the observer. As a result, this source orbital motion, which is referred to as xallarap, is
much stronger for lens systems close to the source, while microlensing parallax signals are stronger when
the lens system is close to the observer. If the xallarap signal is similar to or larger than the parallax signal,
it can generally be distinguished form parallax because light curve models including the xallarap effect will
have a significantly improved $\chi^2$ values, compared to parallax models. This was the case for OGLE-2007-BLG-368
\citep{sumi10}. It is also possible for source companions to influence the photometric signal if the companion
is not much fainter than the source star. For some events, such as MOA-2010-BLG-117 \citep{bennett-moa117},
the effect of the second lensed source is so dramatic that there is no single source, binary lens model that can
provide even an approximate fit to the photometry data. However, it is also possible for a modest magnification
of a companion to the source to slightly perturb the light curve away from the peak in such a way as to perturb
the microlensing parallax signal. This was the case for event MOA-2010-BLG-328, which was published 
with two competing models \citep{moa328}, one with parallax and lens orbital motion, and one with xallarap.
However, the high angular resolution follow-up data from {\sl Keck} and {\sl Hubble} was not consistent with either
of these models. The {\sl Keck} data identified the lens at the expected separation from the source, but there was 
additional stellar flux at the location of the source due to another star. A model including microlensing
parallax, lens star plus planet orbital motion, and the orbital motion and microlensing magnification of
the binary source system, was able to match the positions and magnitudes of the lens and source
stars from the {\sl Keck} and {\sl Hubble} images, as well as the light curve data (Vandorou \etal, in preparation).

In general, the modeling of planetary microlensing events is much simpler when the constraints from
the high angular resolution imaging is included in the modeling, as discussed in Section~\ref{sec-LC}. 
For events events with one-dimensional 
$\piEbold$  measurements, but no other higher order effects, this constrained modeling primarily acts to 
improve the efficiency of MCMC calculations by excluding models that are inconsistent with the
high angular resolution observations from the Markov chains. However, for events like MOA-2010-BLG-328, 
with measurable
orbital motion and/or a source star companion, it becomes more likely that the modeling code 
will be unable to find the correct solution if constraints from the high angular resolution images are not imposed.
The likelihood of modeling failures is increased if the parameters describing the higher order
effects are not constrained to physically reasonable values with a prior distribution.

Another benefit of using the high angular imaging constrains, as discussed in Section~\ref{sec-LC},
is that the constraints can provide redundant measures of the lens system mass and distance.
These redundant measurements 
can then be used to help identify systematic errors in the photometry. This was the case
for event OGLE-2012-BLG-0563 \citep{fukui15}, where systematic
errors in some of the data from microlensing follow-up surveys was found to predict a source radius 
crossing time, $t_*$, that was not consistent with the lens-source separation as measured by both {\sl Keck}
AO and {\sl Hubble} imaging (Bhattacharya \etal; Bennett \etal\ in preparation).

This problem with excluding higher order effects when they don't appear to be necessary to explain the
light curve was foreshadowed in section 6.2 of \citet{penny16}, which notes that six published planets
had published distances of $< 2\,$kpc, when the \citet{penny16} simulations suggested that there should be only
one such planetary microlensing event with the size of the sample they considered. One of these
six events turned out not to have a planetary signal at all \citep{han_ob130723}, and one had a strong
baseline photometry trend that was attributed to the proper motion of nearby star. However, this was
contradicted by more recent data (Udalski, private communication), suggesting that the large microlensing
parallax value, $\pi_{\rm E} \simgt 0.8$, for this event could be spurious.
The remaining four events were part of our program of high angular resolution follow-up imaging for the 22
planetary microlensing event from the \citet{suzuki16} sample. Our analysis indicates that the
previously claimed short lens distances for three of these events, MOA-2007-BLG-192 \citep{bennett08,terry_mb07192}, 
MOA-2010-BLG-328 \citep{moa328}, and OGLE-2012-BLG-0563 \citep{fukui15}, were wrong.

Like the event analyzed in this paper (MOA-2008-BLG-379), MOA-2007-BLG-192 is a high magnification
event with a source magnitude of $I_S \sim 21.5$, but it had a large $\piEbold$ value that turned out 
to be due to a systematic error \citep{koshimoto_blg_pl}, caused by the color dependence of atmospheric 
refraction \citep{bennett12}. This error could have been detected if a Galactic $\piEbold$ prior had been used, 
as this would have revealed that the extremely large $\piEbold$ reported by \citet{bennett08} was highly improbable.


For a robust statistical analysis of a statistical sample of exoplanets found by microlensing, we argue that it is crucial
to include all the higher order microlensing events that can plausibly influence the results, even if the higher order effects
are not well constrained by the data. Otherwise, the values and uncertainties of other microlensing light curve parameters
can be influenced by physically incorrect constraints brought on by setting higher order effect parameters to zero.
It may be necessary to impose prior distributions on some of these higher order effect 
parameters, such as microlensing parallax and lens orbital motion to avoid biasing the results with highly unlikely
values that may be consistent with the data. Note that these prior distributions should be based actual Galactic data.
(Many Bayesian analyses of planetary microlensing events assume that every possible host star has an equal probability 
to host a planet, with the measured mass ratio, $q$, but it is often not recognized that this is a prior assumption that
is not based on any data.) It is also prudent to avoid priors that might be overly prescriptive, as these could interfere
with somewhat unexpected discoveries. Section 6.2 of \citet{penny16}, used a prior assuming that stars in the Galactic
bulge were as likely to host planets and stars in the Galactic disk. This could certainly be considered to
be overly prescriptive. However, the conclusion reached based on this prior provided by this prior was correct. 
The simulations of \citet{penny16} indicated that only one of the planetary systems in their sample should be located
at a distance of $\leq 2\,$kpc, but papers published for 6 of the microlensing planetary systems in his sample indicated
that 6 of these systems were located at distances of $\leq 2\,$kpc. Since then, the distance estimates of 4 of
the claimed 6 planetary systems at $D_L \leq 2\,$kpc have been shown to be wrong, while the distance estimate of one of these
systems is now ambiguous due to a possible systematic photometry error.
We also argue that it is best to impose the constraints from high angular resolution imaging
to the light curve modeling code. This can greatly speed up the analysis by avoiding the exploration of parts of the
parameter space that are inconsistent with the high angular resolution imaging data. For some of the most complicated 
events, these constraints may be necessary to find the correct solutions. However, the possibility that an
overly prescriptive prior might be interfering with the detection of a previously unknown property of planetary systems
must also be considered.

The strengths of this approach are event from our group's lens system mass and distance analysis of the 28 events 
(with 29 planets) in the \citet{suzuki16} statistical sample, (The ``ambiguous event", OGLE-2011-BLG-0950, from 
this sample was found to be due to a stellar binary lens system instead of a planetary system \citep{terry22}.) 
Six of these 28 events had giant source stars, which imply that detecting the exoplanet host 
star is virtually impossible with {\sl Hubble} or {\sl Keck} (although it may be possible with {\sl JWST}), but two of these events
\citep{muraki11,skowron15} have mass measurements from $\pi_{\rm E}$ and $\theta_{\rm E}$ measurements. 
This is also the case for one event, MOA-2010-BLG-117,
which has a source system consisting of a binary pair of subgiants \citep{bennett-moa117}.  
We have obtained high angular resolution follow-up imaging with {\sl Keck} AO or {\sl Hubble} for all 21 events 
without giant or binary sub-giant source stars, and we have detected or 
determined the nature of the lens (and planetary host) star for 12 of them 
\citep{batista15,bennett-ogle109,bennett15,bennett16,bennett_ogle71,aparna18,aparna21,blackman_477,
gaudi-ogle109,terry21,terry_mb07192}, although the publications describing 2 of these events are still
in preparation.

%
%

We have found that image constrained modeling was necessary or useful for 7 of the 12 events with detected host stars,
and for at least 3 of these events (MOA-2008-BLG-379, MOA-2010-BLG-328, and OGLE-2012-BLG-0563), image
constrained modeling was instrumental in obtaining the correct solutions. The modeling of higher order effects was also
crucial for the mass and distance measurements. The three events with giant or binary sub-giant source stars and masses 
from $\piEbold$ measurements (MOA-2009-BLG-266, MOA-2010-BLG-117 and OGLE-2011-BLG-0265) included 
planetary orbital motion in their modeling. Higher order effects, such as additional lens objects, an additional source, or
lens orbital motion were needed for the discovery papers for 
4 of the 12 events with detected or characterized host stars (OGLE-2006-BLG-109,
OGLE-2007-BLG-349, MOA-2010-BLG-117, and MOA-2010-BLG-477). Orbital motion of the lens or source system was
also needed for two of the events without detected host stars, OGLE-2007-BLG-368 \citep{sumi10} and MOA-2009-BLG-387
\citep{batista11}. However, higher order effects beyond those presented in the discovery papers were needed to find the correct
solutions for three of the 12 events with planetary host star detections, MOA-2008-BLG-379 (this paper), MOA-2009-BLG-319
\citep{shin15,terry21}, and MOA-2010-BLG-328 (Vandorou \etal, in preparation). 

We have argued that it is necessary to include higher order effects in microlensing event modeling, even in cases when
microlensing light curves can be fit reasonably well without including these effects. This is because these effects are certain, or
reasonably likely to be present, and ignoring them can lead to errors in the values for the other microlensing light curve parameters.
This might not be considered to be a serious problem, if the analysis 
goal is simply to indicate that the microlensing event is due to a the planetary system, but it
becomes more problematic in a statistical analysis of exoplanetary system properties from microlensing survey, such as the 
one planned for the {\sl Roman Space Telescope} \citep{bennett18_wfirst,penny19,johnson20}. A major advantage
of {\sl Roman's} microlensing survey is its ability to detect the exoplanet host stars and determine the
masses and distances of these systems using the methods that we have used for our
{\sl Keck} adaptive optics and {\sl Hubble} follow-up analysis of planetary microlensing
events observed from the ground. Therefore, we believe that image constrained modeling will be needed to analyze the 
planetary microlensing events found by {\sl Roman}.

\section{Discussion and Conclusions}
\label{sec-conclude}

Our {\sl Keck} AO and {\sl Hubble} follow-up observations have identified the MOA-2008-BLG-379L planetary host 
star through measurements of the host star $K$-band magnitude, the source $V$-band magnitude, the lens and 
source $I$-band magnitudes, and the lens-source relative proper motion, $\mubold_{\rm rel,H}$. 
These measurements constrained some of the light curve parameters and allowed us to determine
host and planet masses and distance through multiple, redundant constraints. We find
host and planet masses of $M_{\rm host} = 0.434\pm 0.065 \msun$, and $m_p = 2.44\pm 0.49 M_{\rm Jup}$,
with a projected separation of $a_\perp = 2.70 \pm 0.42\,$AU at a distance of $D_L = 3.44 \pm 0.53\,$kpc.
These measurements imply that
MOA-2008-BLG-379Lb as the third super-Jupiter mass planet, with a mass in the range 2--3.6$\,M_{\rm Jup}$
orbiting a star of $\sim 0.43\,\msun$ after OGLE-2005-BLG-071Lb \citep{dong-ogle71,bennett_ogle71}
and OGLE-2012-BLG-0406 \citep{poleski_ob120406,tsapras_ob120406}. These discoveries may seem
to disfavor the \cite{laughlin04} argument that gas giants should be rare orbiting M dwarfs, but such a judgement
requires a more detailed statistical analysis. The analysis presented here is part of our campaign to measure masses
for as many of the planets and host stars of the 29 planet complete sample of \cite{suzuki16} as possible. ({\sl Keck} observations 
by \cite{terry22} of the ambiguous event from this sample, favor the stellar binary model over the planetary model.)
We have obtained {\sl Keck} AO observations for all the events in this sample that have source stars with an extinction
corrected source magnitude of $I_{s0} > 16$
under a NASA {\sl Keck} Key Strategic Mission Support program \citep{bennett_KSMS}, and several of the
brighter stars have host and planet mass measurements from a combination of microlensing parallax
measurements and angular Einstein radius determinations from finite source effects
\citep{gaudi-ogle109,bennett-ogle109,muraki11,skowron15,bennett-moa117}.
So, we expect to be able to address this problem more definitively in a future paper that includes
these mass measurements in a statistical analysis.
However, a preliminary statistical analysis including some of these mass measurements does suggest
that the planets found by microlensing, at the measured mass ratios, are more likely to be
hosted by more massive stars. So, perhaps the  \cite{laughlin04} argument does not preclude the hosting 
of super-Jupiter planets by M dwarfs because there is a large dispersion in the properties of protoplanetary
disks, so even though the formation of super-Jupiters may be disfavored around M dwarfs, there are still a 
significant number of M dwarfs that can produce super-Jupiter planets despite this handicap.










\acknowledgments 
The {\sl Keck} Telescope observations and analysis were supported by a NASA {\sl Keck} PI Data Award, administered by the 
NASA Exoplanet Science Institute. Data presented herein were obtained at the W. M. {\sl Keck} Observatory from telescope 
time allocated to the National Aeronautics and Space Administration through the agency's scientific partnership 
with the California Institute of Technology and the University of California. The Observatory was made possible by 
the generous financial support of the W. M. {\sl Keck} Foundation. DPB, AB, NK, SKT, and AV were also supported by 
NASA through grants 80NSSC20K0886 and 80GSFC21M0002. 
Some of this research has made use of the NASA Exoplanet Archive, which is operated by the California Institute of 
Technology, under contract with the National Aeronautics and Space Administration under the Exoplanet Exploration Program.
This work was supported by the University of Tasmania through the UTASFoundation and the endowed Warren Chair 
in Astronomy and the ANR COLD-WORLDS (ANR-18-CE31-0002).
This work was also supported by JSPS Core-to-Core Program JPJSCCA20210003. D.S. was supported by 
JSPS KAKENHI 19KK0082.





\appendix
\section{Light Curve Models Without Parallax with Original and New MOA Photometry}
\label{sec-old-lc}

\begin{deluxetable}{ccccc}
\tablecaption{Close Separation Models with No Microlensing Parallax
                         \label{tab-oldmod} }
\tablewidth{0pt}
\tablehead{
& \multicolumn{3}{c} {2013 MOA photometry} & 2018 MOA phot. \\
\colhead{parameter}  & \colhead{ S14} & \colhead{ S14-new} & \colhead{\texttt{eesunhong}} & \colhead{\texttt{eesunhong}} 
}  
\startdata
$t_{\rm E}$ (days) & $42.46\pm 0.45$ & $42.46\pm 4.93$ & $40.50\pm 6.03$ &  $46.57\pm 5.93$  \\   
$t_0$ (${\rm HJD}^\prime$) & $4687.897\pm 0.001$ & $4687.897\pm 0.001$ & $4687.8937\pm 0.0019$ & $4687.8950\pm 0.0009$   \\
$u_0 \times 10^{3}$ & $ 6.02\pm 0.06$ & $6.02\pm 0.73$ & $ 6.59\pm 1.05$ & $ 5.64\pm 0.79$   \\
$s$ &$ 0.903\pm 0.001$ & $0.903\pm 0.002$ & $0.952\pm 0.011$ & $0.932\pm 0.008$  \\
$\alpha$ (rad) & $1.129\pm 0.002$ & $1.129\pm 0.004$ & $1.1329\pm 0.0035$ & $1.1351\pm 0.0021$  \\
$q \times 10^{3}$ & $6.85\pm 0.05$ & $6.85\pm 0.81$ & $7.04\pm 1.02$ & $6.22\pm 0.78$  \\
$t_\ast$ (days) & $0.0212\pm 0.0020$ & $0.021 \pm 0.003$ &  $0.0208\pm 0.0027$ &  $0.0216\pm 0.0030$  \\
fit $\chi^2$ & 1246.0 & 1246.0 & 1242.80 &1289.96 \\
dof & $\sim 1239$ & $\sim 1239$ & $\sim 1238$ & $\sim 1274$ \\
\enddata
\end{deluxetable}

An early version of this paper noted that the central value of the Einstein radius crossing  
resulting from our image constrained modeling using the \texttt{eesunhong} code ($t_{\rm E} = 55.8\pm 5.5\,$days)
is $\sim 30\sigma$ larger than the best fit value of $t_{\rm E} = 42.46\pm 0.45\,$days quoted in the discovery paper 
\citep{suzuki14,suzuki14e} (for the close model which has a smaller $\chi^2$ value than the wide model). The original version 
of this paper attributed this to the fact that the analysis in the discovery paper did not include microlensing parallax. 
Microlensing parallax was included in the 
initial modeling efforts for this event, but because the source is faint, the odds of detecting a real microlensing parallax
signal were small. The initial analysis did find a formally significant microlensing parallax signal, with a very large 
amplitude that was considered too large to be physical. When the light curve is only sensitive to very large microlensing
parallax amplitudes, it is common for the modeling code to identify modest bumps in the light curve as being caused by
a second approach to the lens. This is because the projected lens-source relative motion is not much larger than the Earth's 
orbital velocity when the microlensing parallax amplitude is large. This occurs regularly when the 
data do not even constrain the microlensing parallax signals to reasonable values, because
the modeling code can probe a large fraction of the light curve
for false parallax signals due to rare systematic photometry errors. \cite{suzuki14,suzuki14e} took the most
common approach when microlensing parallax modeling seems to favor an implausibly large signal caused by systematic
photometry errors: They simply avoided this problem by not including microlensing parallax in the model.

This approach can also be justified by an appeal to Occam's razor, which is often paraphrased as
the statement that ``the simplest explanation is usually the best one\rlap." 
This Occam's razor approach is sensible if one is trying to demonstrate that a specific effect, like microlensing parallax 
or planetary orbital motion, has been detected by the light curve photometry. However, we know the microlensing parallax effect
must exist for Galactic microlensing events observed from a telescope that is being accelerated by the Sun's gravitational
field. Models without microlensing parallax usually yield solutions with physically
reasonable parameters, with one glaring exception: It is unphysical to set ${\bm{\pi_{\rm E}}}\equiv 0$ if the observations were
done with a telescope in orbit about the Sun. In many cases, setting ${\bm{\pi_{\rm E}}}\equiv 0$ will have no practical effect.
However, it is not uncommon for this ${\bm{\pi_{\rm E}}}\equiv 0$ constraint to restrict other light curve parameters, and these
restrictions can exclude the correct models.
This problem with setting ${\bm{\pi_{\rm E}}}\equiv 0$ is evident for 3 microlensing events out of 28 in the 
\citet{suzuki16} statistical 
sample. The $t_{\rm E}$ values for microlensing parallax models differ from the values with ${\bm{\pi_{\rm E}}}\equiv 0$
by $17\sigma$, $12\sigma$, and $6.3\sigma$ for OGLE-2007-BLG-368 \citep{sumi10}, MOA-2009-BLG-387 \citep{batista11},
and OGLE-2012-BLG-0563 \citep{fukui15}, respectively.

A more detailed examination of this issue indicates that the ${\bm{\pi_{\rm E}}}\equiv 0$ is not responsible for the
small $t_{\rm E}$ uncertainty reported in \citet{suzuki14,suzuki14e}. We have remodeled the 2013 photometry used in
the discovery with the modeling code used in the discovery paper and with the \texttt{eesunhong} code, and the results
are compared to the results reported by \citet{suzuki14,suzuki14e} in Table~\ref{tab-oldmod} for the close solutions. The 
best fit models using the code used by \citet{suzuki14,suzuki14e}  are identical, but the error bars are larger for the new
reduction (labeled S14-new) because the new MCMC runs were run long enough to converge. The modeling of the 2013
data with the \texttt{eesunhong} code yield slightly different results due to slightly different treatment of the photometry error
bar estimates. The last column of Table~\ref{tab-oldmod} shows the no-parallax modeling results from the  \texttt{eesunhong} 
code using the 2018 MOA photometry that has systematic errors due to color dependent atmospheric refraction effects
with a detrending method \citep{bennett12,bond17}. Systematic errors due to atmospheric refraction effects are known to 
have an annual correlation that can lead to erroneous microlensing parallax measurements \citep{terry_mb07192}.
Table~\ref{tab-oldmod} shows 
only the close model parameters, but a comparison of the wide models yields
nearly identical results. The analysis of the using the 2013 MOA photometry with the \texttt{eesunhong} modeling 
code excluded  one observation from the baseline as an outlier.
Note that the error bars from the original analysis (labeled S14) are about an order of magnitude too small for
all model parameters except for $t_0$ and $t_*$. The reason for this 
is probably the fact that these parameters
are less affected by the blending degeneracy \citep{yee12}. Evidently, the blending degeneracy was not effectively probed
by the original MCMC analysis.

\vfil\break





\begin{thebibliography}{}
\bibitem[Adams et al.(2021)]{adams21} Adams, F.~C., Meyer, M.~R., \& Adams, A.~D.\ 2021, \apj, 909, 1. doi:10.3847/1538-4357/abdd2b
\bibitem[Alard(1997)]{alard97} Alard, C.\ 1997, \aap, 321, 424 
\bibitem[Alcock et al.(1993)]{macho-lmc1} Alcock, C., Akerlof, C.~W., Allsman, R.~A., et al.\ 1993, \nat, 365, 621. doi:10.1038/365621a0
\bibitem[Alcock et al.(1995)]{macho-par1}Alcock, C., Allsman, R.~A., Alves, D., et al.~1995, \apjl, 454, L125
\bibitem[Adams et al.(2018)]{adams18} Adams, A.~D., Boyajian, T.~S., \& von Braun, K.\ 2018, \mnras, 473, 3608. doi:10.1093/mnras/stx2367
\bibitem[Ali-Dib et al.(2022)]{alidib22} Ali-Dib, M., Cumming, A., \& Lin, D.~N.~C.\ 2022, \mnras, 509, 1413. doi:10.1093/mnras/stab3008
\bibitem[Bachelet et al.(2012)]{bachelet12} Bachelet, E., Fouqu{\'e}, P., Han, C., et al.\ 2012, \aap, 547, A55
\bibitem[Batista et al.(2011)]{batista11} Batista, V., Gould, A., Dieters, S., et al.\ 2011, \aap, 529, A102
\bibitem[Batista et al.(2014)]{batista14} Batista, V., Beaulieu, J.-P., Gould, A., et al.\ 2014, \apj, 780, 54 
\bibitem[Batista et al.(2015)]{batista15} Batista, V., Beaulieu, J.-P., Bennett, D.P., et al.\ 2015, \apj, 808, 170
\bibitem[Beaulieu(2018)]{beaulieu_vvv} Beaulieu, J.-P.\ 2018, Universe, 4, 61. doi:10.3390/universe4040061
\bibitem[Beaulieu et al.(2006)]{ogle390} Beaulieu, J.-P., Bennett, D.~P., Fouqu{\'e}, P., et al.\ 2006, \nat, 439, 437
\bibitem[Beaulieu et al.(2016)]{beaulieu16} Beaulieu, J.-P., Bennett, D.~P., Batista, V., et al.\ 2016, \apj, 824, 83 
\bibitem[Bennett(2008)]{bennett_rev} Bennett, D.P, 2008, in Exoplanets, 
   Edited by John Mason.~Berlin: Springer.~ ISBN: 978-3-540-74007-0,  (arXiv:0902.1761)
\bibitem[Bennett(2010)]{bennett-himag} Bennett, D.P.\ 2010, \apj, 716, 1408
\bibitem[Bennett(2019)]{bennett_KSMS} Bennett, D.\ 2019, Keck Observatory Archive N021, 117
\bibitem[Bennett(2014)]{rhie_obit} Bennett, D.~P.\ 2014, \baas, 46, 008
\bibitem[Bennett et al.(2018a)]{bennett18_wfirst} Bennett, D.~P., Akeson, R., Anderson, J., et al.\ 2018a, (arXiv:1803.08564)
\bibitem[Bennett et al.(2010a)]{bennett_MPF} Bennett, D.~P., Anderson, J., Beaulieu, J.-P., et al.\ 2010a, RFI Response for the Astro2010 decadal survey, arXiv:1012.4486
\bibitem[Bennett et al.(2006)]{bennett06} Bennett, D.~P., Anderson, J., Bond, I.~A., Udalski, A., \& Gould, A.\ 2006, \apjl, 647, L171
\bibitem[Bennett et al.(2007)]{bennett07} Bennett, D.P., Anderson, J., \& Gaudi, B.S.\ 2007, \apj, 660, 781
\bibitem[Bennett et al.(2014)]{bennett14} Bennett, D.~P., Batista, V., Bond, I.~A., et~al.\ 2014, \apj, 785, 155
\bibitem[Bennett et al.(2015)]{bennett15} Bennett, D.~P., Bhattacharya, A., Anderson, J., et al.\ 2015, \apj, 808, 169
\bibitem[Bennett et al.(2020)]{bennett_ogle71} Bennett, D.~P., Bhattacharya, A., Beaulieu, J.-P., et al.\ 2020, \aj, 159, 68. doi:10.3847/1538-3881/ab6212
\bibitem[Bennett et al.(2008)]{bennett08}Bennett, D.~P., Bond, I.~A., Udalski, A., et al.\ 2008, \apj, 684, 663
\bibitem[Bennett \& Khavinson(2014)]{rhie_phystoday} Bennett, D.~P. \& Khavinson, D.\ 2014, Physics Today, 67, 64. doi:10.1063/PT.3.2318
\bibitem[Bennett et al.(2021)]{bennett21} Bennett, D.~P., Ranc, C., \& Fernandes, R.~B.\ 2021, \aj, 162, 243. doi:10.3847/1538-3881/ac2a2b
\bibitem[Bennett \& Rhie(1996)]{bennett96}Bennett, D.P. \& Rhie, S.H.\ 1996, \apj, 472, 660
\bibitem[Bennett \& Rhie(2002)]{bennett02}Bennett, D.P. \& Rhie, S.H.\ 2002, \apj, 574, 985
\bibitem[Bennett et al.(2010)]{bennett-ogle109} Bennett, D.~P., Rhie, S.~H., Nikolaev, S., et~al.\ 2010b, \apj, 713, 837
\bibitem[Bennett et al.(2016)]{bennett16}Bennett, D.P., Rhie, S.H., Udalski, A., et al.\ 2016, \aj, 152, 125
\bibitem[Bennett et al.(2012)]{bennett12} Bennett, D.~P., Sumi, T., Bond, I.~A., et al.\ 2012, \apj, 757, 119 
\bibitem[Bennett et al.(2018)]{bennett-moa117} Bennett, D.~P., Udalski, A., Han, C., et al.\ 2018, \aj, 155, 141 
\bibitem[Bertin \& Arnouts(1996)]{sextractor} Bertin, E., $\&$ Arnouts, S., 1996, A$\&$AS, 117, 393
\bibitem[Bertin et al.(2002)]{SWarp} Bertin, E., Mellier, Y. , Radovich, M.,et al., 2002, The TERAPIX Pipeline, ASP Conference Series, Vol. 281, 228
\bibitem[Bhattacharya et al.(2017)]{aparna17} Bhattacharya, A., Bennett, D.~P., Anderson, J., et al.\ 2017, \aj, 154, 59 
\bibitem[Bhattacharya et al.(2018)]{aparna18} Bhattacharya, A., Beaulieu, J.-P., Bennett, D.~P., et al.\ 2018, \aj, 156, 289 
\bibitem[Bhattacharya et al.(2021)]{aparna21} Bhattacharya, A., Bennett, D.~P., Beaulieu J.~P., et al.\ 2021, \aj, 162, 60
\bibitem[Blackman et al.(2021)]{blackman_477} Blackman, J.~W., Beaulieu, J.~P., Bennett, D.~P., et al.\ 2021, \nat, 598, 272. doi:10.1038/s41586-021-03869-6
\bibitem[Bond et al.(2001)]{bond01} Bond, I.~A., Abe, F., Dodd, R.~J., et al.\ 2001, \mnras, 327, 868
\bibitem[Bond et al.(2017)]{bond17} Bond, I.~A., Bennett, D.~P., Sumi, T., et al.\ 2017, \mnras, 469, 2434. doi:10.1093/mnras/stx1049
\bibitem[Bond et~al.(2004)]{bond04} Bond, I.~A., Udalski, A., Jaroszy{\'n}ski, M.\ 2004,  \apjl, 606, L155
\bibitem[Boyajian et al.(2014)]{boyajian14} Boyajian, T.S., van Belle, G., \& von Braun, K.,\  2014, \aj, 147, 47
\bibitem[Cassan et al.(2012)]{cassan12}Cassan, A., Kubas, D., Beaulieu, J.-P., et al.\ 2012,  \nat, 481, 167
\bibitem[Childs et al.(2022)]{childs22} Childs, A.~C., Martin, R.~G., \& Livio, M.\ 2022, \apjl, 937, L41. doi:10.3847/2041-8213/ac9052
\bibitem[Di Stefano \& Esin(1995)]{distefano95} Di Stefano, R., \& Esin, A.~A.\ 1995, \apjl, 448, L1
\bibitem[Dominik(1999)]{dominik99} Dominik, M.\ 1999, \aap, 349, 108. doi:10.48550/arXiv.astro-ph/9903014
\bibitem[Dominik \& Hirshfeld(1994)]{dominik-lmc1} Dominik, M. \& Hirshfeld, A.~C.\ 1994, \aap, 289, L31
\bibitem[Dong et al.(2009a)]{dong-moa400} Dong, S., Bond, I.~A., Gould, A., et al.\ 2009a, \apj, 698, 1826
\bibitem[Dong et al.(2009b)]{dong-ogle71} Dong, S., Gould, A., Udalski, A., et al.\ 2009b, \apj, 695, 970
\bibitem[Drimmel \& Spergel(2001)]{drimmel}  Drimmel, R., \& Spergel, D.~N.\ 2001, \apj, 556, 181
\bibitem[Emsenhuber et al.(2021)]{emsenhuber21} Emsenhuber, A., Mordasini, C., Burn, R., et al.\ 2021, \aap, 656, A70. doi:10.1051/0004-6361/202038863
\bibitem[Fukui et al.(2015)]{fukui15} Fukui, A., Gould, A., Sumi, T., et al.\ 2015, \apj, 809, 74 
\bibitem[Fulton et al.(2021)]{fulton21} Fulton, B.~J., Rosenthal, L.~J., Hirsch, L.~A., et al.\ 2021, \apjs, 255, 14. doi:10.3847/1538-4365/abfcc1
\bibitem[Furusawa et al.(2013)]{moa328} Furusawa, K., Udalski, A., Sumi, T., et al.\ 2013, \apj, 779, 91
\bibitem[Gaudi(2012)]{gaudi_araa} Gaudi, B.~S.\ 2012, \araa, 50, 411
\bibitem[Gaudi et al.(2008)]{gaudi-ogle109} Gaudi, B.~S., Bennett, D.~P., Udalski, A., et al.\ 2008, Science, 319, 927
\bibitem[Gaudi \& Gould(1997)]{gaudi97} Gaudi, B.S., \& Gould, A.\ 1997, \apj, 486, 85
\bibitem[Ghosh et al.(2004)]{ghosh04} Ghosh, H., DePoy, D.~L., Gal-Yam, A., et al.\ 2004, \apj, 615, 450
\bibitem[Gonzalez et al.(2011)]{vvv_extinct} Gonzalez, O.~A., Rejkuba, M., Zoccali, M., Valenti, E., \& Minniti, D.\ 2011, \aap, 534, A3
\bibitem[Gould(1992)]{gould-par1} Gould, A.\ 1992, \apj, 392, 442
\bibitem[Gould(2014)]{gould-1dpar} Gould, A.\ 2014, J.\ Kor.\ Ast.\ Soc., 47, 215
\bibitem[Gould et al.(2010)]{gould10}Gould, A., Dong, S., Gaudi, B.S., et al.\ 2010,  \apj, 720, 1073
\bibitem[Gould \& Loeb(1992)]{gouldloeb92} Gould, A. \& Loeb, A. 1992, \apj, 396, 104
\bibitem[Gould et al.(1994)]{gould-diskorhalo}  Gould, A., Miralda-Escude, J., \& Bahcall, J.~N.\ 1994, \apjl, 423, L105
\bibitem[Gould et al.(2006)]{gould06} Gould, A., Udalski, A., An, D., et al.\ 2006, \apjl, 644, L37
\bibitem[Gould et al.(2009)]{gould09} Gould, A., Udalski, A., Monard, B., et al.\ 2009, \apjl, 698, L147
\bibitem[Grazier(2016)]{grazier16} Grazier, K.~R.\ 2016, Astrobiology, 16, 23. doi:10.1089/ast.2015.1321
\bibitem[Griest \& Safizadeh(1998)]{griest98} Griest, K., \& Safizadeh, N.\ 1998, \apj, 500, 37
\bibitem[Han et al.(2016)]{han_ob130723} Han, C., Bennett, D.~P., Udalski, A., \& Jung, Y.~K.\ 2016, \apj, in press (arXiv:1604.06533)
\bibitem[Holtzman et al.(1998)]{holtzman98} Holtzman, J.~A., Watson, A.~M., Baum, W.~A., et al.\ 1998, \aj, 115, 1946
\bibitem[Ida \& Lin(2004)]{idalin04} Ida, S.\ \& Lin, D.N.C.\ 2004, \apj, 604, 388
\bibitem[Janczak et al.(2010)]{janczak10} Janczak, J., Fukui, A., Dong, S., et al.\ 2010, \apj, 711, 731
\bibitem[Johnson et al.(2010)]{johnson10} Johnson, J.~A., Aller, K.~M., Howard, A.~W., \& Crepp, J.~R.\ 2010, \pasp, 122, 905 
\bibitem[Johnson et al.(2020)]{johnson20} Johnson, S.~A., Penny, M., Gaudi, B.~S., et al.\ 2020, \aj, 160, 123. doi:10.3847/1538-3881/aba75b
\bibitem[Kervella et al.(2004)]{kervella_dwarf} Kervella, P., Th{\'e}venin, F., Di Folco, E., \& S{\'e}gransan, D.\ 2004, \aap, 426, 297
\bibitem[Koshimoto et al.(2021a)]{koshimoto_gal_mod} Koshimoto, N., Baba, J., \& Bennett, D.~P.\ 2021a, \apj, 917, 78. doi:10.3847/1538-4357/ac07a8
\bibitem[Koshimoto et al.(2020)]{koshimoto_bayes_follow} Koshimoto, N., Bennett, D.~P., \& Suzuki, D.\ 2020, \aj, 159, 268. doi:10.3847/1538-3881/ab8adf
\bibitem[Koshimoto et al.(2021b)]{koshimoto_blg_pl} Koshimoto, N., Bennett, D.~P., Suzuki, D., et al.\ 2021b, \apjl, 918, L8. doi:10.3847/2041-8213/ac17ec
\bibitem[Laughlin et al.(2004)]{laughlin04} Laughlin, G.  Bodenheimer, P.\ \& Adams, F.C.\ 2004, \apjl, 612, L73
\bibitem[Lecavelier des Etangsa \& Lissauer(2022)]{exoplanet_IAU} Lecavelier des Etangsa, Jack J. Lissauer, J.~J.\ 2022, arXiv:2203.09520
\bibitem[Lissauer(1993)]{lissauer_araa} Lissauer, J.J.\ 1993, Ann.\ Rev.\ Astron.\ Ast., 31, 129
\bibitem[Lu (2008)]{jlu_thesis} Lu, J.~R., \ 2008,  Exploring the Origins of Young Stars in the Central Parsec of our Galaxy
with Stellar Dynamics, UCLA Ph.D. Thesis
\bibitem[Lu (2022)]{KAI} Lu, J., Keck-Data Reduction Pipelines/KAI: v1.0.0 Release of KAI (v1.0.0). Zenodo. https://doi.org/10.5281/zenodo.6522913
\bibitem[Mao \& Paczynski(1991)]{mao91} Mao, S. \& Paczynski, B.\ 1991, \apjl, 374, L37. doi:10.1086/186066
\bibitem[Mayor et al.(2011)]{mayor11} Mayor, M., Marmier, M., Lovis, C., et al.\ 2011, arXiv:1109.2497
\bibitem[Minniti et al.(2010)]{minniti-vvv} Minniti, D., Lucas, P.~W., Emerson, J.~P., et al.\ 2010, \na, 15, 433 
\bibitem[Mulders et al.(2015)]{mulders15} Mulders, G.~D., Pascucci, I., \& Apai, D.\ 2015, \apj, 798, 112. doi:10.1088/0004-637X/798/2/112
\bibitem[Mordasini et al.(2009)]{mordasini09} Mordasini, C., Alibert, Y., \& Benz, W.\ 2009, \aap, 501, 1139
\bibitem[Muraki et al.(2011)]{muraki11} Muraki, Y., Han, C., Bennett, D.~P., et al.\ 2011, \apj, 741, 22
\bibitem[Nataf et al.(2013)]{nataf13} Nataf, D.~M., Gould, A., Fouqu{\'e}, P., et al.\ 2013, \apj, 769, 88 
\bibitem[Nayakshin et al.(2019)]{nayakshin19} Nayakshin, S., Dipierro, G., \& Szul{\'a}gyi, J.\ 2019, \mnras, 488, L12. doi:10.1093/mnrasl/slz087
\bibitem[Nayakshin et al.(2022)]{nayakshin22} Nayakshin, S., Elbakyan, V., \& Rosotti, G.\ 2022, \mnras, 512, 6038. doi:10.1093/mnras/stac833
\bibitem[Nishiyama et al.(2006)]{nish06} Nishiyama, S., Nagata, T., Kusakabe, N., et al.\ 2006, \apj, 638, 839. doi:10.1086/499038
\bibitem[Osinski et al.(2020)]{osinski20} Osinski, G.~R., Cockell, C.~S., Pontefract, A., et al.\ 2020, Astrobiology, 20, 1121. doi:10.1089/ast.2019.2203
\bibitem[Penny et al.(2019)]{penny19} Penny, M.~T., Gaudi, B.~S., Kerins, E., et al.\ 2019, \apjs, 241, 3
\bibitem[Penny et al.(2016)]{penny16} Penny, M.~T., Henderson, C.~B., \& Clanton, C.\ 2016, \apj, 830, 150. doi:10.3847/0004-637X/830/2/150
\bibitem[Poleski et al.(2014)]{poleski_ob120406} Poleski, R., Udalski, A., Dong, S., et al.\ 2014, The Astrophysical Journal, 782, 47
\bibitem[Pollack et al.(1996)]{pollack96} Pollack, J.~B., Hubickyj, O., Bodenheimer, P., et al.\ 1996, \icarus, 124, 62
\bibitem[Poindexter et al.(2005)]{poindexter05} Poindexter, S., Afonso, C., Bennett, D.~P., et al.\ 2005, \apj, 633, 914. doi:10.1086/468182
\bibitem[Quenoille (1949)]{quenouille1949} Quenouille, M.~H. 1949, The Annals of Mathematical Statistics, 20, 355
\bibitem[Quenoille (1956)]{quenouille1956} Quenouille, M.~H. 1956, Biometrika. 43, 353
\bibitem[Rattenbury et al.(2017)]{rattenbury17} Rattenbury, N.~J., Bennett, D.~P., Sumi, T., et al.\ 2017, \mnras, 466, 2710 
\bibitem[Raymond et al.(2004)]{raymond04} Raymond, S.~N., Quinn, T., \& Lunine, J.~I.\ 2004, \icarus, 168, 1. doi:10.1016/j.icarus.2003.11.019
\bibitem[Raymond et al.(2007)]{raymond07} Raymond, S.~N., Quinn, T., \& Lunine, J.~I.\ 2007, Astrobiology, 7, 66. doi:10.1089/ast.2006.06-0126
\bibitem[Rhie et al.(1999)]{rhie_98smc1} Rhie, S.~H., Becker, A.~C., Bennett, D.~P., et al.\ 1999, \apj, 522, 1037
\bibitem[Rhie \& Bennett(1996)]{rhie-lmc1} Rhie, S.~H. \& Bennett, D.~P.\ 1996, Nuclear Physics B Proceedings Supplements, Vol. 51, 51, 86. doi:10.1016/S0920-5632(96)00487-2

\bibitem[Rhie et al.(2000)]{rhie00} Rhie, S.~H., Bennett, D.~P., Becker, A.~C., et al.\ 2000, \apj, 533, 378
\bibitem[Rosenthal et al.(2021)]{rosenthal21} Rosenthal, L.~J., Fulton, B.~J., Hirsch, L.~A., et al.\ 2021, \apjs, 255, 8. doi:10.3847/1538-4365/abe23c
\bibitem[Schlecker et al.(2022)]{schlecker22} Schlecker, M., Burn, R., Sabotta, S., et al.\ 2022, \aap, 664, A180. doi:10.1051/0004-6361/202142543
\bibitem[Service et al.(2016)]{distortion} Service, M., Lu, J.~R., Campbell, R., et al.\ 2016, \pasp, 128, 095004
\bibitem[Shin et al.(2015)]{shin15} Shin, I.-G., Han, C., Choi, J.-Y., et al.\ 2015, \apj, 802, 108. doi:10.1088/0004-637X/802/2/108
\bibitem[Sinclair et al.(2020)]{sinclair20} Sinclair, C.~A., Wyatt, M.~C., Morbidelli, A., et al.\ 2020, \mnras, 499, 5334. doi:10.1093/mnras/staa3210
\bibitem[Skowron et al.(2015)]{skowron15} Skowron, J., Shin, I.-G., Udalski, A., et al.\ 2015, \apj, 804, 33. doi:10.1088/0004-637X/804/1/33
\bibitem[Spergel et al.(2015)]{WFIRST_AFTA} Spergel, D., Gehrels, N., Baltay, C., et al.\ 2015, arXiv:1503.03757 
\bibitem[Stetson(1987)]{Daophot} Stetson, P.~B., \ 1987, PASP, 99, 191S
\bibitem[Sumi et al.(2010)]{sumi10}Sumi, T., Bennett, D.~P., Bond, I.~A. et al.\ 2010,  \apj, 710, 1641
\bibitem[Sumi et al.(2016)]{sumi16} Sumi, T., Udalski, A., Bennett, D.~P., et al.\ 2016, \apj, 825, 112 
\bibitem[Surot et al.(2020)]{surot20} Surot, F., Valenti, E., Gonzalez, O.~A., et al.\ 2020, \aap, 644, A140. doi:10.1051/0004-6361/202038346
\bibitem[Suzuki et al.(2018)]{suzuki18} Suzuki, D., Bennett, D.~P., Ida, S., et al.\ 2018, \apjl, 869, L34
\bibitem[Suzuki et al.(2016)]{suzuki16} Suzuki, D., Bennett, D.~P., Sumi, T., et al.\ 2016, \apj, 833, 145
\bibitem[Suzuki et al.(2014)]{suzuki14} Suzuki, D., Udalski, A., Sumi, T., et al.\ 2014, \apj, 780, 123 
\bibitem[Suzuki et al.(2014e)]{suzuki14e} Suzuki, D., Udalski, A., Sumi, T., et al.\ 2014e, \apj, 788, 97 
\bibitem[Szul{\'a}gyi et al.(2014)]{szulagyi14} Szul{\'a}gyi, J., Morbidelli, A., Crida, A., et al.\ 2014, \apj, 782, 65. doi:10.1088/0004-637X/782/2/65
\bibitem[Szyma{\'n}ski et al.(2011)]{ogle3-phot} Szyma{\'n}ski, M.~K., Udalski, A., Soszy{\'n}ski, I., et al.\ 2011, \actaa, 61, 83 
\bibitem[Terry et al.(2024)]{terry_mb07192} Terry, S.~K., Beaulieu, J.-P., Bennett, D.~P., et al.\ 2024, arXiv:2403.12118. doi:10.48550/arXiv.2403.12118
\bibitem[Terry et al.(2022)]{terry22} Terry, S.~K., Bennett, D.~P., Bhattacharya, A., et al.\ 2022, \aj, 164, 217. doi:10.3847/1538-3881/ac9518
\bibitem[Terry et al.(2021)]{terry21} Terry, S.~K., Bhattacharya, A., Bennett, D.~P., et al.\ 2021, \aj, 161, 54. doi:10.3847/1538-3881/abcc60
\bibitem[Tierney \& Mira (1999)]{tierney} Tierney, L., $\&$ Mira, A. 1999, Stat Med, 18, 2507
\bibitem[Thompson et al.(2018)]{kepler2018} Thompson, S.~E., Coughlin, J.~L., Hoffman, K., et al.\ 2018, \apjs, 235, 38
\bibitem[Tsapras et al.(2014)]{tsapras_ob120406} Tsapras, Y., Choi, J.-Y., Street, R.~A., et al.\ 2014, \apj, 782, 48
\bibitem[Tukey (1958)]{tukey1958} Tukey, J. W. 1958, The Annals of Mathematical Statistics, 29, 614
\bibitem[Udalski et al.(2005)]{udalski05} Udalski, A., Jaroszy{\'n}ski, M., Paczy{\'n}ski, B., et al.\ 2005, \apjl, 628, L109
\bibitem[Udalski et al.(1994)]{ogle-ews} Udalski, A., Szyma\'{n}ski, M., Ka{\l}u\.{z}ny, J., Kubiak, M., Mateo, M., Krzmi\'{n}ski, W., \& \pac, B.\ 1994, Acta Astron., 44, 227
\bibitem[Udalski et al.(2015)]{udalski_ogle124} Udalski, A., Yee, J.~C., Gould, A., et al.\ 2015, \apj, 799, 237. doi:10.1088/0004-637X/799/2/237
\bibitem[Vandorou et al.(2020)]{van20} Vandorou, A., Bennett, D.~P., Beaulieu, J.-P., et al.\ 2020, \aj, 160, 121. doi:10.3847/1538-3881/aba2d3
\bibitem[Yee et al.(2012)]{yee12} Yee, J.~C., Shvartzvald, Y., Gal-Yam, A., et al.\ 2012, \apj, 755, 102
\bibitem[Zhang et al.(2022)]{zhang22} Zhang, K., Gaudi, B.~S., \& Bloom, J.~S.\ 2022, Nature Astronomy, 6, 782. doi:10.1038/s41550-022-01671-6
\end{thebibliography}
\end{document}